\documentclass{aa}  

\usepackage[english]{babel}
\usepackage{graphicx}	
\usepackage{amsmath}	
\usepackage[utf8]{inputenc} 
\usepackage{newlfont} 
\usepackage{indentfirst}
\usepackage{subcaption}
\usepackage{float}
\usepackage{wrapfig}
\usepackage{amssymb}
\usepackage{mathtools}
\usepackage{physics}
\usepackage{latexsym}
\usepackage{dirtytalk} 
\usepackage[table]{xcolor}
\usepackage{fancyhdr}
\usepackage{tabularx}
\usepackage{multirow}
\usepackage{array}
\usepackage{siunitx}
\usepackage{comment}
\usepackage{listings}
\usepackage{xcolor}
\usepackage{stfloats}
\usepackage{txfonts}
\usepackage[breaklinks=true]{hyperref} 

\hypersetup{
    colorlinks=true,
    linkcolor=blue,
    urlcolor=blue,
    citecolor=blue,
}

\newcommand{\HeII}{\textrm{He\,\textsc{ii}}}
\newcommand{\NV}{\textrm{N\,\textsc{v}}}
\newcommand{\Halpha}{H$\alpha$}
\newcommand{\Hbeta}{H$\beta$}
\newcommand{\OIII}{\textrm{O\,\textsc{iii}}}

\DeclareSIUnit\parsec{pc}
\DeclareSIUnit\kiloparsec{kpc}
\DeclareSIUnit\megaparsec{Mpc}
\DeclareSIUnit\solarmass{M_{\odot}}
\DeclareSIUnit\year{yr}
\DeclareSIUnit\elettronvolt{eV}
\DeclareSIUnit\dex{dex}
\DeclareSIUnit\h{h}
\DeclareSIUnit\angstrom{\mathring{A}}

\definecolor{codegreen}{rgb}{0,0.6,0}
\definecolor{codegrey}{rgb}{0.5,0.5,0.5}
\definecolor{codepurple}{rgb}{0.58,0,0.82}
\definecolor{backcolour}{rgb}{0.95,0.95,0.92}

\lstdefinestyle{mystyle}{
    backgroundcolor=\color{backcolour},   
    commentstyle=\color{codegreen},
    keywordstyle=\color{magenta},
    numberstyle=\tiny\color{codegrey},
    stringstyle=\color{codepurple},
    basicstyle=\ttfamily\footnotesize,
    breakatwhitespace=false,         
    breaklines=true,                 
    captionpos=b,                    
    keepspaces=true,                 
    numbers=left,                    
    numbersep=5pt,                  
    showspaces=false,                
    showstringspaces=false,
    showtabs=false,                  
    tabsize=2
}

\begin{document} 

   \title{Probing the dawn of galaxies: star formation and feedback in the JWST era through the GAEA model}

   \author{Sebastiano Cantarella \inst{1,2}\fnmsep\thanks{\email{sebastiano.cantarella@phd.units.it}} \and Gabriella De Lucia\inst{2,3} \and Fabio Fontanot\inst{2,3} \and Michaela Hirschmann\inst{4,2} \and Lizhi Xie\inst{5} \and Maximilien Franco\inst{6} \and Adèle Plat\inst{4}}

   \institute{Astronomy Section, Department of Physics, University of Trieste, via G.B. Tiepolo 11, I-34143, Trieste, Italy
   \and 
   INAF -- Astronomical Observatory of Trieste, via G.B. Tiepolo 11, I-34143 Trieste, Italy
   \and
   IFPU -- Institute for Fundamental Physics of the Universe, Via Beirut 2, 34151 Trieste, Italy
   \and 
   Institute of Physics, GalSpec Laboratory, EPFL, Observatoire de Sauverny, Chemin Pegasi 51, CH-1290 Versoix, Switzerland
   \and
    Tianjin Normal University, Binshuixidao 393, Tianjin, China
   \and
    Université Paris-Saclay, Université Paris Cité, CEA, CNRS, AIM, 91191 Gif-sur-Yvette, France}

   \date{Received MM, DD, YYYY; accepted MM, DD, YYYY}
 
  \abstract{The James Webb Space Telecope (JWST) opened a new window for the study of the highest redshift ($z > 7$) Universe. This work presents a theoretical investigation of the very-high redshift Universe using the state-of-the-art GALaxy Evolution and Assembly (GAEA) model, run on merger trees from the Planck-Millennium $N$-body simulation. We show that GAEA successfully reproduces a wide range of high-$z$ observational estimates including: the galaxy stellar mass function up to $z \sim 13$ and the total (galaxies and AGN) UV luminosity function (LF) up to $z \sim 10$. We find that the AGN UV emission represents an important contribution at the bright end of the UVLF up to $z \sim 8$, but it is negligible at higher redshift. Our model reproduces well the observed mass-metallicity relation at $z \leq 4$, while it slightly overestimates the normalization of the relation at earlier cosmic epochs. At $z \geq 11$, current UVLF estimates are at least one order of magnitude larger than model predictions. We investigate the impact of different physical mechanisms, such as an enhanced star formation efficiency coupled with a reduced stellar feedback or a negligible stellar feedback at $z > 10$. In the framework of our model, both the galaxy stellar mass and UV luminosity functions at $z \geq 10$ can be explained by assuming feedback-free starbursts in high-density molecular clouds. However, we show that this model variant leads to a slight increase of the normalization of the $z \geq 10$ mass-metallicity relation, strengthening the tension with available data. A model with negligible stellar feedback at $z > 10$ also predicts larger numbers of massive and bright galaxies aligning well with observations, but it also overestimates the metallicity of the interstellar medium. We show that these model variants can in principle be discriminated using the relation between the star formation rate and galaxy stellar mass.}

   \keywords{galaxies: formation -- galaxies: evolution -- galaxies: high-redshift -- galaxies: stellar content -- galaxies: luminosity function -- galaxies: abundances}

  \authorrunning{Cantarella, S.}
  \titlerunning{High-redshift galaxies with GAEA in the JWST era}
   
   \maketitle

\section{Introduction}\label{sec:introduction}

The first three years of observations from the James Webb Space Telescope (JWST) have yielded groundbreaking discoveries for the very-high redshift Universe. Early results from surveys conducted using NIRCam have broken the `redshift barrier' at $z \sim 8-10$ established by pre-JWST observations, detecting very massive and bright galaxies within the Epoch of Reionization and beyond \citep[e.g.,][]{donnan2023a, donnan2023b, labbe2023, harikane2023a, finkelstein2023}. Subsequent spectroscopic follow-ups have shown that some of these photometric candidates were actually interlopers at lower redshifts \citep[e.g.,][]{arrabalharo2023}, leading initially to some overestimation of the number density of massive galaxies at early epochs. However, numerous galaxies have been confirmed spectroscopically to be beyond $z \sim 10$ \citep[e.g.,][]{harikane2023b, carniani2024, napolitano2025, naidu2025}. Although its relatively small field of view makes it challenging to survey large cosmic volumes, JWST's unique combination of sensitivity and infrared wavelength coverage is revolutionizing our view of the very-high-redshift Universe. For the first time, it is providing strong constraints on the faint end of the galaxy stellar mass function (GSMF) and on the rest-frame ultraviolet luminosity function (UVLF), deep into the Epoch of Reionization. 

The GSMF has been massively studied in the literature, from the local Universe up to very early cosmic epochs \citep[e.g.,][]{muzzin2013,stefanon2021,weaver2023,harvey2025}. Studies at high redshift rely mainly on rest-frame optical photometric data and are based on imaging from both ground and space-based telescopes. Recent observations of the $z \geq 6$ Universe using JWST have been carried out for regions of the sky in dedicated surveys such as CEERS \citep[\SI{35.5}{arcmin}$^2$]{finkelstein2023} and JADES \citep[\SI{220}{arcmin}$^2$]{bunker2020}. These surveys have revealed the presence of UV luminous galaxies (possibly massive, according to photometric estimates) at very-high redshifts \cite[e.g.,][]{robertson2023b,curtis-lake2023,robertson2023b} corresponding to  \SIrange{300}{500}{\mega\year} after the Big Bang. The number densities estimated for these massive galaxies challenge current theoretical models \citep[e.g.,][]{wilkins2022,finkelstein2023,finkelstein2024,yung2024a}. The small sky coverage of these surveys means that their results are susceptible to cosmic variance, which could help reconcile the apparent discrepancy with theoretical models \citep[e.g.,][]{mccaffrey2023}. In addition, it is important to bear in mind that current estimates of the GSMF still carry significant uncertainties (e.g., photometric redshift uncertainties, uncertainties associated with the SED fitting procedure, etc.).

Important insights are provided by the rest-frame UVLF that provides information about the star formation activity, the growth of supermassive black holes (SMBHs) over cosmic time, and dust physics. Before the launch of JWST, the UVLF at $z \gtrsim 6$ has been studied through a combination of ground-based near-infrared surveys \citep[e.g., UltraVISTA,][]{mccracken2012, bowler2020} and space-based observations with the Hubble Space Telescope, Spitzer and Swift \citep[e.g.,][]{finkelstein2015,hagen2015,parsa2016,bouwens2021a}. The advent of JWST has provided observational evidence that the UVLF continues to evolve beyond $z \sim 10$, in contrast to what simple constant star formation efficiency models predict \citep[e.g.,][]{bouwens2023a,bouwens2023b,harikane2023a,finkelstein2025}. The observed UVLF is found to be in agreement with theoretical predictions  up to $z \sim 10$, while observational measurements are larger than published model predictions at earlier cosmic epochs \citep[e.g.,][]{yung2024a}. 

The current situation is that measurements based on JWST data point to an excess of both massive and UV-bright galaxies at $z \gtrsim 10$ compared to theoretical predictions. A plethora of possible explanations has been proposed: observational estimates might be affected by a selection bias against the oldest and least star forming galaxies \citep{mason_trenti_treu2023, shen2023}; radiatively driven outflows may evacuate the dust produced by stars in the early stages of galaxy build-up, enhancing its observed UV luminosity \citep{fiore2023,ziparo2023}; UV luminosities might be boosted by a top-heavy initial mass function (IMF) associated with Pop III stars or by AGN activity \citep{harikane2023a,finkelstein2023,yung2024a,ferrara2024,ferrara2025}; the star formation efficiency (SFE) at $z > 10$ might be enhanced due to a lack of effective feedback \citep{harikane2022a,dekel2023}. Other possible explanations have been explored outside the context of the current $\Lambda$CDM model, such as a change of the power spectrum \citep{padmanabhan_loeb2023c} or alternative cosmologies \citep[e.g.,][]{melia2023,liu2024}. Different physical prescriptions have been tested, showing that the contribution of dust attenuation is non-negligible up to $z \sim 11$ \citep{mauerhofer2025}, that either a top-heavy or an evolving IMF \citep{trinca2024,yung2024a}, along with an evolving SFE \citep{mauerhofer2025, somerville2025}, can alleviate tensions between model predictions and observations, and that galaxies dominated by Pop III stars could explain the UV-bright excess at $z \geq 12$ \citep{ventura2024}.

All the physical processes invoked to explain the overabundance of massive and UV‐bright systems at $z \gtrsim 10$ also regulate the production and retention of metals in young galaxies. Thus, the same feedback and inflow/outflow cycles that shape the GSMF and UVLF must leave a distinct signature in the mass–metallicity relation (MZR). Mapping the MZR from the local Universe \citep[e.g.,][]{tremonti2004} up to the Epoch of Reionization  \citep[e.g., recent JWST/NIRSpec measurements at $4 \leq z \leq 9$;][]{nakajima2023,morishita2024} therefore provides a crucial, physically independent test for galaxy formation models. None of the studies mentioned above have simultaneously considered these independent constraints. 

In this work, we use the GAlaxy Evolution and Assembly (GAEA) model to study the physics of $z \gtrsim 4$ galaxies and to investigate whether a self-consistent physical picture can explain the available observations. Specifically, we study the contribution of AGN and the impact of dust on the UV luminosities, and explore the consequences of different star formation and feedback efficiencies on the observed GSMF and UVLF at $z \geq 10$, and on the predicted mass-metallicity relation.

This paper is organized as follows. Section \ref{sec:info_PMS_and_GAEA} introduces the galaxy formation model and the simulations used in our study. Sections \ref{sec:GAEA_fiducial_GSMF}, \ref{sec:GAEA_fiducial_UVLF} and \ref{sec:GAEA_fiducial_MZR} compare model predictions with observational estimates, focusing in particular on: the UV luminosity function, the galaxy stellar mass function, the stellar-to-halo mass ratio, and the mass-metallicity relation. In Section \ref{sec:model_variants} we discuss our results in a broader framework considering a number of model variants and studying their impact on high-$z$ model predictions. Section \ref{sec:summary_conclusion} summarises our findings and gives our conclusions.

\section{Simulations and galaxy formation model}\label{sec:info_PMS_and_GAEA}

In this work, we take advantage of a state-of-the-art galaxy formation model, based on a semi-analytic approach. The backbone of semi-analytic models (SAMs) is a statistical representation of the assembly of dark matter (DM) halos (i.e., a `merger tree'). The latter can be either built analytically or extracted from a numerical realization of a statistically representative volume of the Universe, using $N$-body or Lagrangian simulations. SAMs then populate DM halos with baryons and trace the formation and evolution of galaxies by numerically solving a series of coupled differential equations that describe the physical mechanisms governing the evolution of the baryonic components. These prescriptions typically include a set of free parameters that reflect our limited understanding of the physical processes involved, and that are fine-tuned to reproduce a specific subset of observational constraints in the local Universe. The approach is limited by the lack of spatially resolved information for modelled galaxies and of an explicit treatment for gas dynamics. However, a key advantage is represented by the significantly reduced computational costs for evolving galaxies within large cosmological volumes. Therefore, SAMs provide a convenient and powerful tool for statistical studies of galaxy evolution, enabling testing various physical prescriptions and their impact on observed galaxy properties.

\subsection{Planck Millennium N-body simulation}\label{sec:P-Millennium_Nbody_details}

The Planck Millennium simulation \citep[hereafter PMS;][]{baugh2019} has been run on a memory reduced version of the $N$-body code GADGET-2 \citep{springel2005e} and follows the evolution of $5040^3$ DM particles within a cosmological box of \SI{542.16}{\megaparsec\per\h} on a side, with initial conditions generated at $z = 127$. The six cosmological parameters ($\Omega_m$, $\Omega_b$, $\Omega_\Lambda$, $h$, $n$ and $\sigma_8$) adopted in the PMS are consistent with Planck14 data \citep{planck2014}. They differ from the ones adopted in the previous simulations of the suite (the Millennium and MillenniumII Simulations - hereafter MS and MSII, respectively), that are instead consistent with the first-year results from WMAP \citep[WMAP1; see][]{spergel2003}. The combination of both a large number of particles and a large cosmological box size for the PMS leads to a resolution that is intermediate between MS and MSII: the latter simulations use $2160^3$ particles in cosmological boxes of size \SI{500}{\megaparsec\per\h} and \SI{100}{\megaparsec\per\h}, respectively. As for the MS and MSII, halos and subhalos were identified using a Friends-of-Friends (FoF) algorithm and the SUBFIND code \citep{2001MNRAS.328..726S}. Substructures with at least $20$ bound particles are retained as genuine substructures, that corresponds to a halo mass resolution limit of $2.12 \times 10^{9}$ \SI{}{\solarmass\per\h} for the PMS. Subhalo-based merger trees are built using the same approach described in \cite{springel2005c} -- see also \cite{fontanot2025}.

Halos and subhalos information for the PMS is stored at $272$ snapshots, a significantly larger number of outputs than that used for the MS and MSII ($64$ and $68$, respectively). In particular, the PMS has a finer sampling of the cosmic evolution of DM haloes at high redshift, with $\sim 100$ snapshots at $z \geq 4$, compared to the $\sim 25$ outputs available for the MS and MSII, making it ideal for studying high redshifts.

In the following sections, we will present model predictions based on merger trees that have been created using half of the available snapshots from the PMS. In Appendix \ref{app:compare_PMS_runs} we show that model outputs are not affected by the adopted sampling.

\subsection{GAEA}\label{sec:GAEA_details}

In this work, we take advantage of the latest version of the GAEA theoretical model of galaxy formation and evolution \citep{delucia2024}. GAEA builds on the original model presented in \cite{delucia_blaizot2007} but it has been updated significantly over time. In particular, the current rendition of the model includes:
\begin{itemize}
    \item a treatment for chemical enrichment that includes a non-instantaneous recycling of gas, metals, and energy \citep{delucia2014};
    \item  a parametrization of stellar feedback based partially on results of high-resolution hydro-dynamical simulations \citep{hirschmann2016}, and resulting in a good agreement with the observed low-mass end of the galaxy stellar mass function up to $z \sim 4$;
    \item an explicit partitioning of cold gas into its atomic and molecular components \citep{xie2017};
    \item a detailed tracing of angular momentum exchanges between different components, and a model for the non-instantaneous stripping of cold and hot gas in satellite galaxies \citep{xie2020};
    \item improved prescriptions for cold gas accretion onto supermassive black holes (SMBHs) and outflows from Active Galactic Nuclei (AGN) \citep{fontanot2020b}.
\end{itemize}

The latest version of GAEA, as its previous renditions, has been calibrated using the $z \leq 3$ GSMF, the HI and H$_2$ galaxy mass functions at $z = 0$, and the $z < 4$ AGN luminosity function. In this work, we adopt the same values of the model parameters used in \citet{delucia2024} and in \cite{fontanot2025}, and assume a universal \cite{chabrier2003} IMF. The only modification we introduce is relative to the prescription for metal ejection after stellar feedback: in previous renditions of our model, we have assumed that for haloes less massive than  $\hat{M}_{vir} = 3 \times 10^{10}$ \SI{}{\solarmass} all newly synthesized metals would be injected directly into the hot gaseous phase \citep{delucia2014}. In this paper, we assume that all newly formed metals are ejected into the inter-stellar medium of galaxies, independently of the parent halo mass. The original prescription was introduced to better reproduce the metal abundance of Milky-Way satellites. Therefore, our modification does not significantly affect predictions for more massive galaxies. We discuss the impact of this modification in more detail in Section \ref{sec:GAEA_fiducial_MZR}.

The black hole seeding mechanism currently adopted in GAEA has been introduced in \cite{xie2017} and is based on a scaling relation between the mass of the black hole and that of the parent halo. The growth of the SMBH is assumed to be driven by both galaxy mergers and disc instabilities, as described in \cite{fontanot2020b}.

As discussed in previous papers, our model reproduces a variety of observational measurements over a broad range of cosmic epochs including: the galaxy stellar mass function up to $z \sim 7$ and the cosmic star formation rate density up to $z \sim 10$ \citep{hirschmann2016,fontanot2017b}; the galaxy mass-metallicity relation and its secondary dependence on the star formation rate and the gas mass \citep{delucia2020}; the observed evolution of the galaxy mass – gas/star metallicity relations \citep{fontanot2021}; the evolution of the fraction and number densities of quenched galaxies up to $z \sim 4$ \citep{delucia2024}; and the strength of the galaxy clustering signal as a function of stellar mass and star formation rate up to $z \sim 3.5$ \citep{fontanot2025}. Such agreement is preserved in the version of GAEA used in this paper.

\section{The UV Luminosity Function of galaxies and AGN}\label{sec:GAEA_fiducial_UVLF}

The UV luminosity of a galaxy is produced mostly by young, massive stars (it is a tracer of recent star formation) and AGN (the UV luminosity traces the accretion of gas onto the disk around a SMBH). Distinguishing these contributions is not trivial from the observational point of view, and requires spectroscopic information. In particular, deep spectroscopy of the host galaxy is needed to assess the presence of an AGN through their characteristic emission lines. Type-$1$ AGN can be identified via broad permitted lines (like \Halpha, \Hbeta, \HeII,  etc.), and Type-$2$ AGN via the appearance of high-ionization lines such as \NV, and also via specific line-ratios \citep[see, e.g.,][]{hirschmann2019,hirschmann2023b,hirschmann2023a}. However, direct UV spectroscopic measurements of AGN are not always available, thus the contribution to UV luminosity due to  AGN has to be derived through conversions from other wavelength bands (e.g., X-ray). X-ray selected AGN samples suffer from spectroscopic incompleteness as well as non-negligible uncertainties on the intrinsic physical properties of the AGN and on error propagation due to the luminosity conversion \citep{lusso_risaliti2016, kulkarni_worseck_hennawi2019}.

In the last decade, there have been several attempts to disentangle the contribution of AGN and stars to the observed UV luminosity of individual objects \citep[e.g.,][]{akiyama2018,finkelstein_bagley2022,harikane2022a}. Thanks to spectroscopic information coming from JWST and other telescopes (e.g., Subaru), it has recently become possible to better constrain the number density of AGN at early cosmic epochs \citep{harikane2022a}. Available estimates are affected by non-negligible systematics due to assumptions on the priors, and to uncertainties related to the method adopted to derive the LF that become non-negligible for small-area surveys \citep{weaver2022}. 

On the theoretical side, estimating the contribution to UV luminosity from stars and AGN requires some assumptions. In this paper, we use the absolute magnitude $M_{UV}$ in the AB system \citep{oke_gunn1983}, and assume a top-hat filter centred on $\SI{1600}{\angstrom}$. For each galaxy, $M_{UV}$ is computed using the  stellar population models from \cite{bruzual_charlot2003}. To estimate dust attenuation, we adopt the prescriptions described in \cite{delucia_blaizot2007} based on the original model proposed by \cite{charlot_fall2000}.

\begin{figure*}
    \centering
    \includegraphics[width=0.95\linewidth]{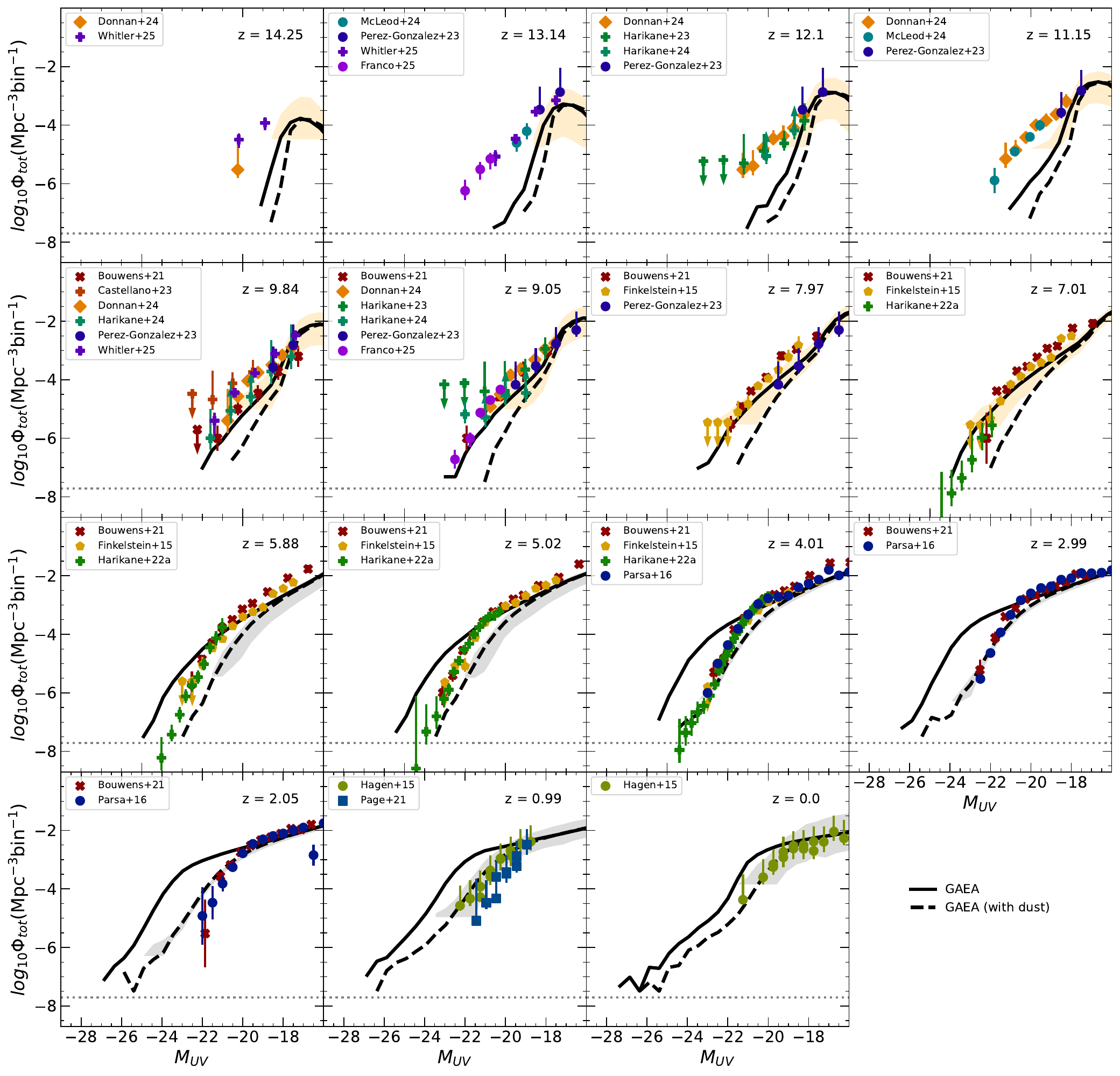}
    \caption{Rest-frame galaxy UV Luminosity Function at $0\leq z \leq 14$ from GAEA, compared with observational estimates \citep{finkelstein2015,hagen2015,parsa2016, bouwens2021a,page2021,castellano2023,perez-gonzalez2023,harikane2022a,harikane2023a,harikane2024a,donnan2024,mcleod2024, franco2025, whitler2025}. The solid and dotted lines represent model predictions with and without  dust attenuation, respectively. The grey horizontal lines show the space density corresponding to 10 sources in the PMS volume. The grey and orange shaded areas show the effect of cosmic variance on the dust-attenuated and the unattenuated UVLF, respectively (see text for details). Different symbols refer to different techniques for computing the LF: circle for the $1 / V_\mathrm{max}$ method, plus sign for the effective volume approach, cross sign for the STY method, diamond for the Bayesian approach, pentagon for the SWML method, square for the binned estimation of the LF. Arrows in downward direction are referring to upper limits, while arrows in upward direction to lower limits.}
    \label{fig:GAEA_galUVLF_fiducial}
\end{figure*}

\begin{figure*}[h!]
    \centering
    \includegraphics[width=0.95\linewidth]{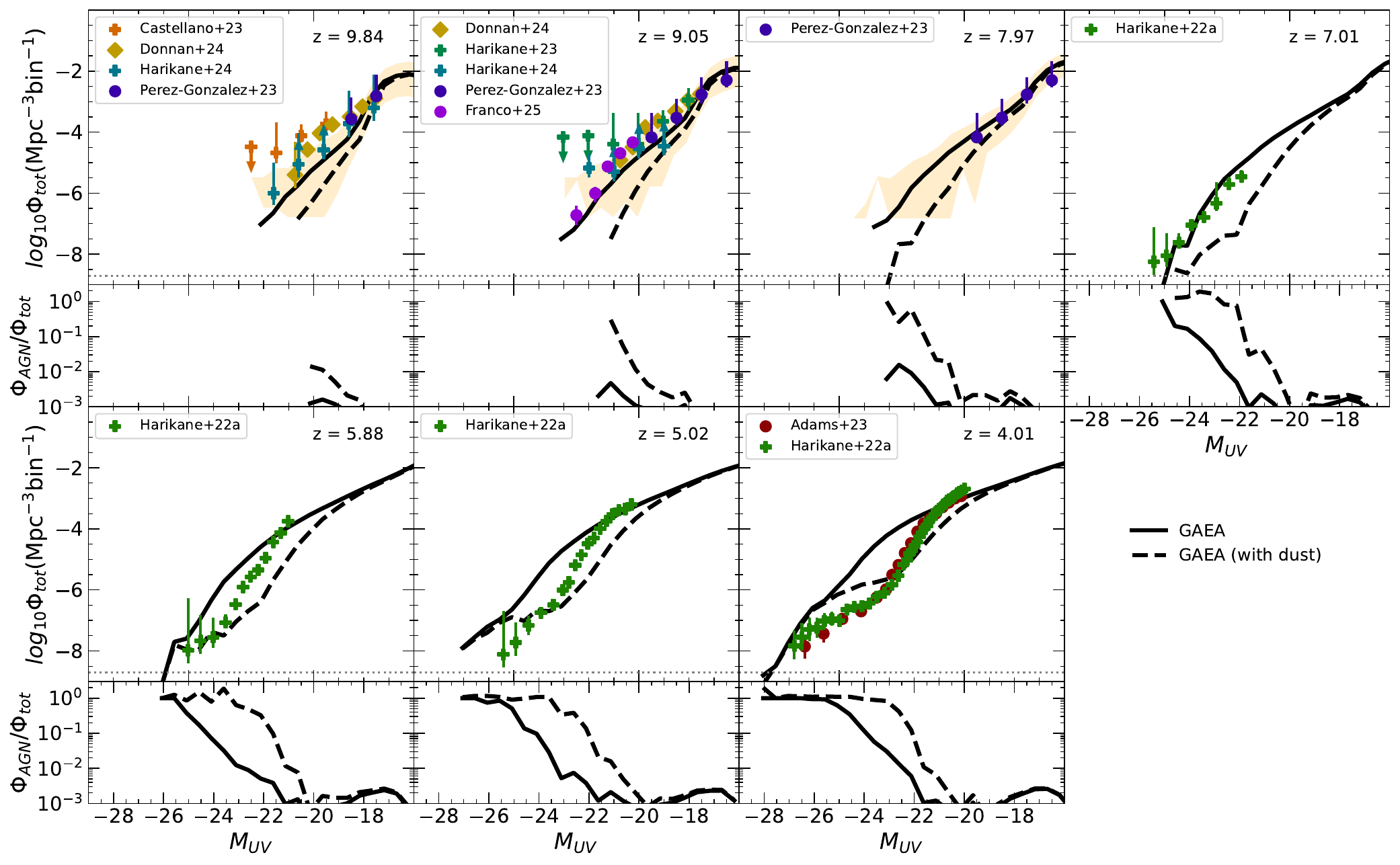}
    \caption{Total (galaxies + AGN) rest-frame UV luminosity function at $3 \leq z \leq 9$ as predicted by GAEA, compared to observational data \citep{adams2023b,castellano2023,perez-gonzalez2023,donnan2024,harikane2022a,harikane2023a,harikane2024a}. Symbols and lines are as in \autoref{fig:GAEA_galUVLF_fiducial}. The sub-panels show the ratio between the AGN UVLF and the total UVLF at each redshift, with (solid) and without (dashed) dust attenuation. The orange shaded areas show the effect of cosmic variance on the dust-unattenuated UVLF (see text for details).}
    \label{fig:GAEA_totUVLF_fiducial}
\end{figure*}

To compute the UV luminosity of an AGN, we adopted two different methods. In the first one, we start computing the bolometric luminosity as
\begin{equation}\label{eq:bol_lum_AGN}
    L_{\mathrm{bol}} = \frac{\epsilon_{\mathrm{rad}}}{1 - \epsilon_{\mathrm{rad}}} \dot {M}_{\mathrm{BH}} c^2 \ ,
\end{equation}
where $\epsilon_{\mathrm{rad}} = 0.15$ is the assumed radiative efficiency, and $\dot{M}_{\mathrm{BH}}$ represents the mass accretion rate on the corresponding SMBH. $\dot{M}_{\mathrm{BH}}$ is an output of the model that is computed according to the accretion rate prescriptions described in \cite{fontanot2020b}. We then estimate the B-band luminosity $L_\mathrm{B}$ adopting the bolometric correction from \cite{marconi2004} (Equation $21$, third line). To convert the B-band into UV-band luminosity, we assume a power law \begin{equation}\label{eq:power_law}
    L(\lambda) \propto \lambda^{-\alpha}\ ,
\end{equation}
with $\alpha = 0.44$ and reference wavelengths $\lambda_\mathrm{B} = \SI{4400}{\angstrom}$ and $\lambda_\mathrm{UV} = \SI{1600}{\angstrom}$. We note that this is the method used for calibrating GAEA with the UVLF of AGN.

Alternatively, we use the Eddington ratio $f_\textup{Edd}$ of the SMBH and its mass $M_\textup{BH}$ and compute the AGN UV luminosity adopting the models of \cite{kubota_done2019} for super Eddington accretion (slim disc model), and from \cite{kubota_done2018} for sub-Eddington accretion (thin disc). In the sub-Eddington case, we use the models `\textsc{qsosed}’, that assume an empirical relation between $f_\textup{Edd}$ and the X-ray luminosity. This relation is valid down to $\log_{10}(f_\textup{Edd}) = -1.5$; below this value we use `\textsc{agnsed}’ models with default parameters \citep[see Appendix~A in][]{kubota_done2018}. We note that the super Eddington slim disc models are only available up to $\log(M_\textup{BH}) = 10$, so above this value we use the thin disc accretion model. 

In both cases, we do not apply any obscuration correction for the AGN luminosity. The two approaches give consistent results but at the bright end of the total (galaxy + AGN) rest-frame UVLF, where the second method predicts lower number densities with respect to the first method. We show such comparison in Appendix \ref{app:comp_AGN_UVLF_techniques}, while in the following we will show only predictions coming from the second approach. 

\autoref{fig:GAEA_galUVLF_fiducial} shows the predicted rest-frame UV luminosity function of galaxies at $0 \leq z \leq 14$. Solid and dashed lines refer to UVLF obtained considering the intrinsic and dust obscured UV magnitudes, respectively. Model predictions are compared with different observational estimates. Although the UVLF estimators are not the same for the considered UVLFs (see the caption of \autoref{fig:GAEA_galUVLF_fiducial}), they are consistent within the uncertainties, especially at $z \leq 8$.

In each sub-panel, the grey dotted horizontal line shows the number densities corresponding to $10$ galaxies in the PMS volume. The impact of cosmic variance on the dust-attenuated $z \leq 7$ UVLF is shown by the grey solid lines, while the orange solid lines show the contribution of cosmic variance on the unattenuated $z > 7$ UVLF. To estimate the cosmic variance, we consider independent subvolumes corresponding to a projected area equal to the survey sky area considered at each redshift and a depth corresponding to the redshift range of the observations. We then tile the PMS volume with the maximum number of non-overlapping subvolumes corresponding to the same effective volumes of the data. Specifically, we assume the following reference sky areas: \cite{whitler2025} ($160$ arcmin$^2$) at $z = 14$, \cite{perez-gonzalez2023} ($75$ arcmin$^2$) at $9 \leq z \leq 13$, \cite{finkelstein2015} ($301.2$ arcmin$^2$) at $5 \leq z \leq 8$, \cite{parsa2016} ($0.292$ deg$^2$) at $2 \leq z \leq 4$ and \cite{hagen2015} ($\sim270$ arcmin$^2$) at $0 \leq z \leq1$.

At $z \leq 4$, GAEA model predictions including dust attenuation are in quite good agreement with observational estimates. At $5 \leq z \leq 6$ the observed UVLFs appear to require  less dust attenuation than that predicted from the model at $M_{UV} \lesssim -20$. Moreover, the estimated impact of cosmic variance at $5 \leq z \leq 10$ introduces a $\sim \SIrange{0.5}{0.75}{\dex}$ scatter in the intrinsic predictions at every redshift, reducing the offset between observations and dust-attenuated predictions. At $z > 7$, observational estimates of the UVLF are better described by the intrinsic luminosities of model galaxies. Therefore, to reproduce available data, GAEA requires a marked decrease of dust attenuation at $ z> 7$, in agreement with expectations based on the limited time available for dust growth \citep[see e.g.,][]{mauerhofer2025}. However, this conclusion appears to be in contrast with recent work pointing to a significant impact of dust attenuation on the total star formation rate density (and therefore UV luminosity) at  $ 7.5 \lesssim z \lesssim 9$ \citep{martis2025}. Our current modelling of dust attenuation is simplistic and calibrated to observations in the local Universe. Therefore, we do not deem our predictions for dust obscuration to be reliable at high redshifts. An explicit implementation of dust formation, growth and destruction is in progress and  will be presented in a forthcoming work (Osman et al., in preparation).

At $z \geq 10$, the model underestimates the number density of galaxies for $M_{UV} \lesssim -18$. In this redshift range, the impact of cosmic variance does not help to explain the mismatch between the model and the data, and this tension is shared by other recently published galaxy formation models \citep[e.g.,][]{yung2024a}. Several alternative explanations have been proposed. 

One possibility is a non-negligible contribution from AGN to the $z \gtrsim 10$ UVLF. We address this possibility in \autoref{fig:GAEA_totUVLF_fiducial}, showing model predictions for the total rest-frame UVLF at $4 \leq z \leq 10$. We highlight the AGN contribution to the total UVLF by computing $\Phi_\mathrm{AGN} / \Phi_\mathrm{tot}$ and showing it in each sub-panel. At $4\leq z \leq 5$, an AGN contribution is required to reproduce the bright end of the observed total UVLF, although our model predictions overestimate the UV-bright LF. At these redshifts, the contribution from the cosmic variance (computed using the sky areas from \cite{perez-gonzalez2023} at $8 \leq z \leq 10$ and \cite{harikane2023a} ($\sim $\SI{300}{\deg^2}) at $4 \leq z \leq 7$) is negligible. At $7 \leq z \leq 9$, the predicted UVLFs fall within the uncertainty range allowed by cosmic variance and the model can reproduce the observed UVLFs if dust attenuation is less important than predicted from our current rendition. The AGN contribution to the bright end of the UVLF decreases as redshift increases, in agreement with observational estimates \citep{finkelstein_bagley2022}. At $z \geq 7$, the observed UVLF is in quite good agreement with our model predictions with no dust attenuation, and the AGN contribution is predicted to be significant only at the very bright end. In conclusion, we find that in the framework of our model the $3 \leq z \leq 8$ UVLF is AGN-dominated -- i.e., with $\Phi_\mathrm{AGN} / \Phi_\mathrm{tot} \geq 10\%$ -- at $M_{UV} \lesssim -23$, while above $z \sim 9$ the entire UV range observed is galaxy-dominated.

These results agree well with independent theoretical models. For example, \cite{trinca2024} used the CAT galaxy formation model, which is calibrated at $z = 4$ and does not provide predictions below that redshift, to study the total UVLF at $4 \leq z \leq 16$. The authors found that the UVLF is AGN-dominated at $M_{UV} \lesssim -22$ up to $z \sim 7$ and that AGN contribute only $\lesssim 1$ percent to the total UVLF at $7 \leq z \leq 10$. At $z \geq 10$, they find that the AGN contribution is negligible and the UVLF is galaxy-dominated.

Another possible explanation for the lack of UV luminous galaxies at $z \gtrsim 10$ in model predictions might be related to the assumed star formation efficiency and stellar feedback. A weaker stellar feedback would allow more gas to cool and convert into new stars, enhancing the UV luminosity. A larger star formation efficiency would also lead to an enhanced production of massive stars, although this would in turn increase the effect of stellar feedback, i.e. the self-regulation between these two processes could balance in such a way to leave the predicted UVLF almost unaffected. In Section \ref{sec:model_variants}, we discuss in more detail the impact of these modifications on model predictions. 

A top-heavy IMF at high redshift could also lead to an enhancement of the abundance of UV-bright galaxies \citep{finkelstein2023,harikane2024a,yung2024a}, reconciling model predictions with observations \citep{harikane2023a}. This could be motivated by considering that the inter-stellar medium of high redshift galaxies has a lower metallicity, favouring the formation of more massive stars \citep{chon2022}. \cite{yung2024a} evaluated the impact of a `top-heavy IMF' by manually boosting the UVLF \textit{a posteriori}. This `boost' translates into a rigid shift along the $M_{UV}$ axis towards brighter luminosities. The authors argue that a boost of a factor $\sim 4$ could bring their model predictions in agreement with observational data. Using an independent galaxy formation model, \cite{mauerhofer2025} studied the evolution of the UVLF assuming an IMF that depends on both redshift and metallicity (`eIMF' in their work). They find that, while both modifications lead to an enhancement of the number densities of bright UV galaxies, model predictions assuming an evolving IMF can be hardly distinguished from those from a model that assumes a universal IMF and an evolving star formation efficiency (`eSFE' in their work) depending on halo mass and redshift. Our model includes a self-consistent treatment for a varying IMF \citep{fontanot2017a,fontanot2018a,fontanot2024a}. We postpone to a future work a detailed study of the impact of a variable IMF scenario,  also in the context of a detailed accounting of the impact of Pop III stellar populations.

\section{The galaxy stellar mass function of high-redshift galaxies}\label{sec:GAEA_fiducial_GSMF}

\begin{figure*}[h!]
    \centering
    \includegraphics[width=0.95\linewidth]{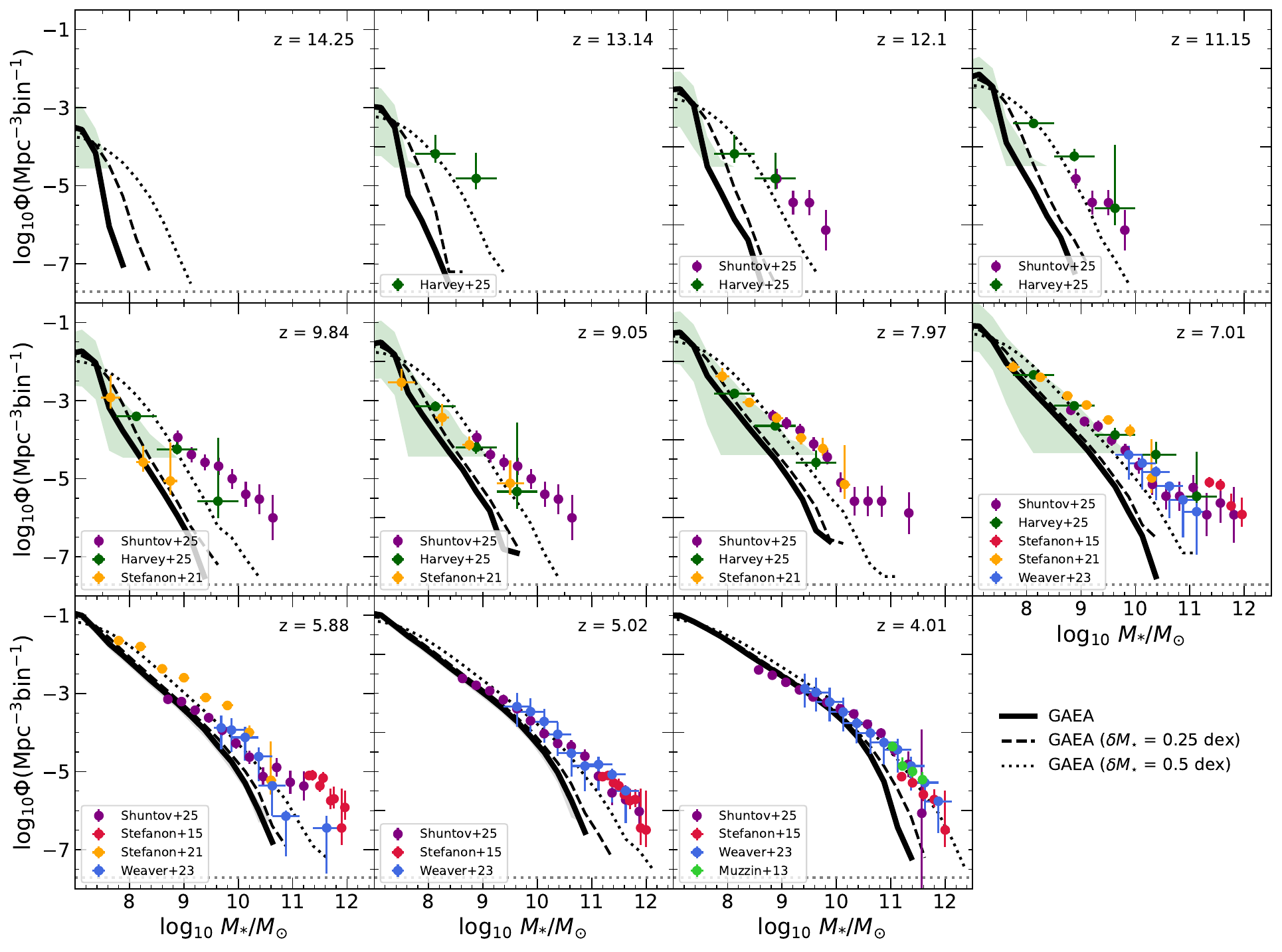}
    \caption{The galaxy stellar mass function at $4 \leq z \leq 14$ from the GAEA semi-analytical model used in this study. Symbols with associated error bars correspond to the observational measurements from \cite{harvey2025} for $7 < z < 13$, \cite{stefanon2015} for $4 < z < 7$, \cite{stefanon2021} for $6 < z < 10$, \cite{weaver2023} for $4 < z < 7$, \cite{muzzin2013} for $z = 4$, and \cite{shuntov2025} for $4 \leq z \leq 12$. Solid lines represent the intrinsic model predictions, while dashed and dotted lines are obtained assuming an uncertainty on stellar mass of  \SI{0.25}{\dex} and \SI{0.5}{\dex}, respectively. The green and grey shaded areas show the impact of cosmic variance on the intrinsic model predictions when survey sky areas from \cite{harvey2025} and the COSMOS2025 field are considered, respectively. The dotted horizontal grey lines mark the number densities corresponding to $10$ galaxies within the simulated volume.}
    \label{fig:GAEA_GSMF_fiducial}
\end{figure*} 

As mentioned in Section \ref{sec:GAEA_details}, GAEA reproduces the observed GSMF at $z \leq 3$ by construction. \cite{fontanot2025} have shown that predictions from the same model are robust against changes in the resolution and small changes in cosmological parameters of the backbone simulation -- in particular, they have compared predictions from merger trees extracted from the MS, MSII and PMS. In the following, we focus only on model predictions based on the PMS and on a redshift range ($z \geq 4$) that goes beyond the calibration set adopted and that we have largely unexplored up to now. 

Figure \ref{fig:GAEA_GSMF_fiducial} shows the predicted $4 \leq z \leq 14$ GSMFs from GAEA run on the PMS. Solid lines show the intrinsic model predictions. All observational estimates of the GSMFs shown in Figure \ref{fig:GAEA_GSMF_fiducial} have been obtained using the $1 / V_{max}$ method \citep{schmidt1968}. We note that model predictions discussed in this work are computed at specific snapshots ($z = 4.01$, $5.02$, $5.88$, $7.01$, $7.97$, $9.05$, $9.84$, $11.15$, $12.10$, $13.14$, $14.25$), while observational estimates typically correspond to relatively wide redshift ranges. We have verified that computing the GSMF combining snapshots around different redshift intervals ($\Delta z = 0.1$ up to $z = 5$; $\Delta z = 0.2$ at $5 < z < 10$; $\Delta z = 0.5$ at $z \geq 10$) leads to an increase in the number density at the high-mass end, due to the larger effective volume. These GSMF estimates are in fact between the GSMF obtained from individual snapshots when assuming an observational uncertainty on galaxy stellar mass in the range of \SIrange{0.25}{0.5}{\dex} (see dashed and dotted lines in the figure). Therefore, we do not show them in Figure \ref{fig:GAEA_GSMF_fiducial}.

The GSMF evolves significantly at redshift $z \geq 7$, both in terms of stellar mass and number density. At all redshifts shown, the intrinsic model predictions are systematically below the observed GSMFs, with an offset at the most massive end that appears to increase with increasing redshift. As well understood, the Eddington bias plays an important role in the comparison between data and models. In particular, because of the typical exponential cutoff of the GSMF at the high-mass end, the Eddington bias makes massive galaxies up-scatter in more massive bins, translating into larger number densities for the most massive galaxies. We take this into account by showing model predictions obtained assuming an uncertainty of $0.25$ and \SI{0.5}{\dex} (dashed and dotted lines in Figure \ref{fig:GAEA_GSMF_fiducial}, respectively). At $z \sim 4$, an observational uncertainty of the order \SI{0.25}{\dex} is enough to bring model predictions in rather good agreement with observational estimates. At higher redshifts, model predictions come close to observational measurements only assuming a larger uncertainty, of the order of \SI{0.5}{\dex}. These are not unrealistic considering uncertainties on photometric redshifts and SED fitting (which includes assumptions about the stellar IMF and the star formation histories). The impact of Poisson noise and cosmic variance also becomes important at high redshift and for the most massive galaxies \citep[e.g., see Fig.~$3$ in][]{shuntov2025}.

Regarding the various observational estimates considered, we note that the data by \cite{stefanon2015} (red circles) are systematically above other data at $z = 6 - 7$ for galaxies with $\log_{10}(M_* / M_\odot) \geq 11$. This discrepancy is primarily driven by systematic effects in the derivation of photometric redshifts, that are related to the adoption of SED templates for old and dusty galaxies and to bayesian priors on the observed flux. Moreover, as mentioned earlier, the small number of massive galaxies in the sample and the significant impact of cosmic variance introduce large statistical uncertainties. Together, these factors limit the reliability of any constraint on the evolution of the high-mass end of the GSMF across the redshift range $4 < z < 7$. 

At $z \gtrsim 8$, measurements from \cite{shuntov2025} (purple circles) are also based on a single field of \SI{0.53}{\deg\squared} and might be significantly affected by increasing uncertainties on photometric redshifts, stellar mass estimates and sample incompleteness. 

To quantify the impact of cosmic variance on the predicted GSMF, we assume two survey areas: a combination of PEARLS and other JWST surveys ($187$ arcmin$^2$) for $z \geq 7$ (as used by \cite{harvey2025}), and COSMOS2025 (\SI{0.54}{\deg\squared}) for $z < 7$ (as used by \cite{shuntov2025}). We follow the same procedure described in Section \ref{sec:GAEA_fiducial_GSMF}. Results are shown in \autoref{fig:GAEA_GSMF_fiducial} as green and grey shaded regions for the fields from \cite{harvey2025} and COSMOS2025, respectively. At $z > 7$, cosmic variance introduces a scatter of $\SIrange{0.5}{1}{\dex}$  increasing towards larger stellar masses, so that the intrinsic and the perturbed GSMFs at $z \geq 7$ become consistent with observational data.

\begin{figure*}[h!]
    \centering
    \includegraphics[width=0.95\linewidth]{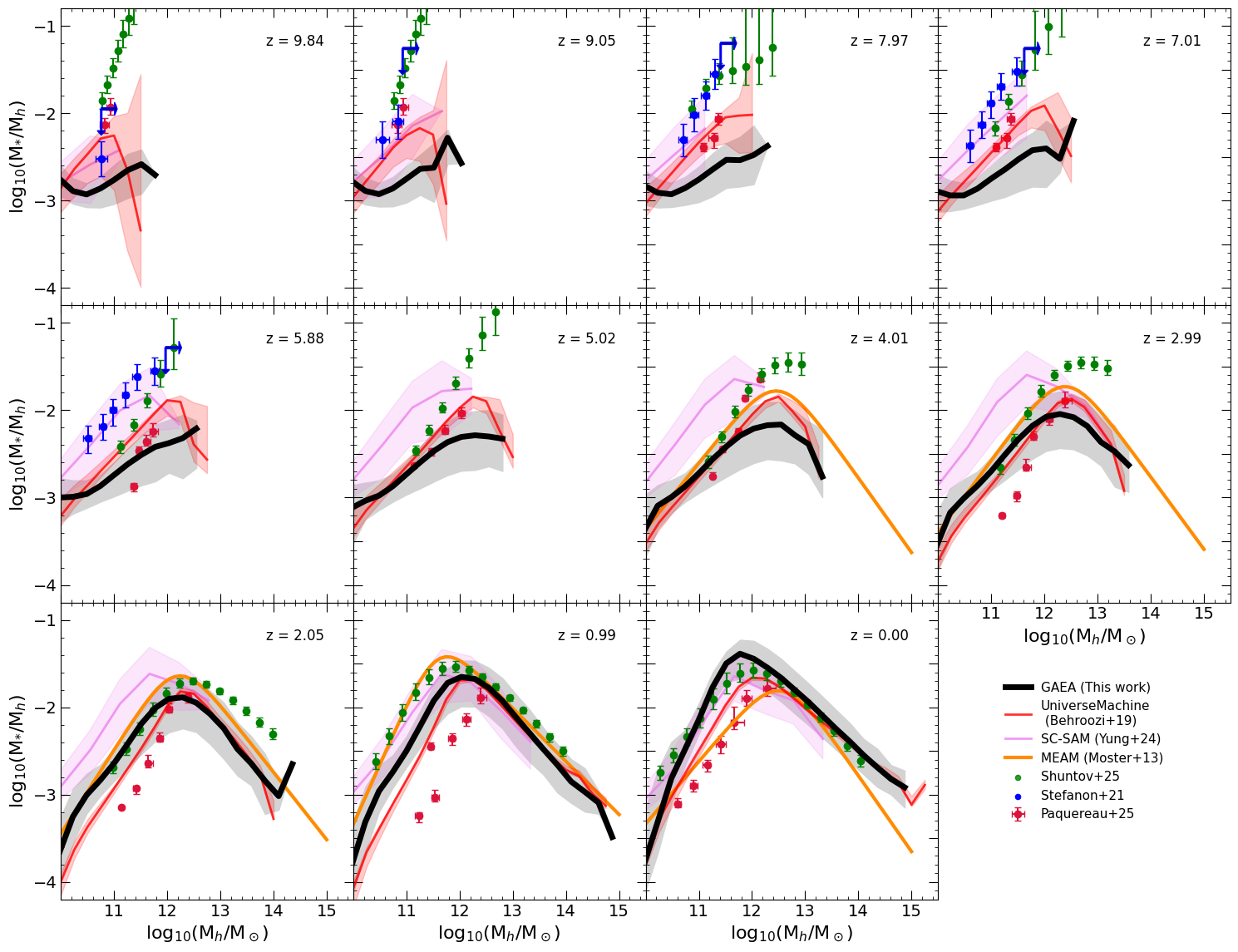}
    \caption{Stellar-to-Halo Mass Relation at $0 \leq z \leq 10$ from GAEA, compared with observational estimates \citep[point with error bars]{stefanon2021,shuntov2025,paquereau2025} and theoretical work (lines with shaded areas). We include both empirical abundance matching models (MEAM from \cite{moster2013}; UniverseMachine from \cite{behroozi2019}) and the Santa-Cruz SAM \citep{yung2024a}. The solid lines represent the median relations (the black ones refer to GAEA), while the shaded regions show the $1 \sigma$ confidence regions.}
    \label{fig:GAEA_SHMR_fiducial}
\end{figure*}

Another way to look at the GSMF is studying the integrated star formation efficiency through the stellar-to-halo mass relation (hereafter SHMR). \autoref{fig:GAEA_SHMR_fiducial} shows predictions from GAEA along with observational measurements and independent theoretical predictions. The observational estimates by \cite{stefanon2021} and \cite{shuntov2025} are obtained employing an abundance matching method (AM) that assumes that each halo hosts one galaxy. Measurements from \cite{paquereau2025} adopt a standard halo occupation distribution (HOD) approach, in which the number of galaxies inside a halo is assumed to depend only on its mass $M_h$. Our predictions agree very well with results of \cite{moster2013} and \cite{behroozi2019}, considering the $1 \sigma$ uncertainties, up to $z = 4$. On the other hand, the normalization of the relation from \cite{yung2024a}, that is based on the Santa-Cruz semi-analytic model, is systematically higher than ours, although still consistent within the scatter. In addition, these predictions are close to the ones from \cite{behroozi2019} at $z > 4$. This implies that in these two models the stellar mass growth of galaxies is likely to be substantial at earlier cosmic epochs -- i.e., it requires a higher integrated SFE at $z \gtrsim 10$ than in GAEA. 

It is worth noting that the observational estimates from \cite{shuntov2025} and \cite{paquereau2025} are based on the same data set \citep[Cosmos-Web]{casey2023} but exhibit a discrepancy. This difference stems from the different galaxy-halo connection approach employed. In detail, the relation shown in \cite{paquereau2025} has been computed from the ratio $M_{*,th} / M_{h,min}$, where $M_{h,min}$ is the characteristic halo mass for which $50$ percent of halos host at least one central galaxy, and $M_{*,th}$ is the minimum observed stellar mass considered in the analysis. Because this relation is built from threshold values, it is driven by the low-mass end of the stellar mass function of central galaxies. By contrast, the $M_* / M_h$ estimates reported by \cite{shuntov2025} (and by \cite{stefanon2021}) represent median relations derived via abundance matching and are therefore mainly driven by the massive end of the stellar mass distribution of central galaxies. Therefore, the relation provided by \cite{paquereau2025} shown here can be treated as an observational lower limit within the context of the SHMR.

Our model predictions are in very good agreement with estimates of \cite{shuntov2025} and with all the theoretical and empirical relations reported here within the $1 \sigma$ confidence level, up to $z \sim 1$. At higher redshifts, estimates by \cite{shuntov2025}and \cite{stefanon2021} lie systematically above theoretical and empirical expectations.

\section{Mass-Metallicity Relation}\label{sec:GAEA_fiducial_MZR}

\begin{figure*}
    \centering
    \includegraphics[width=0.95\linewidth]{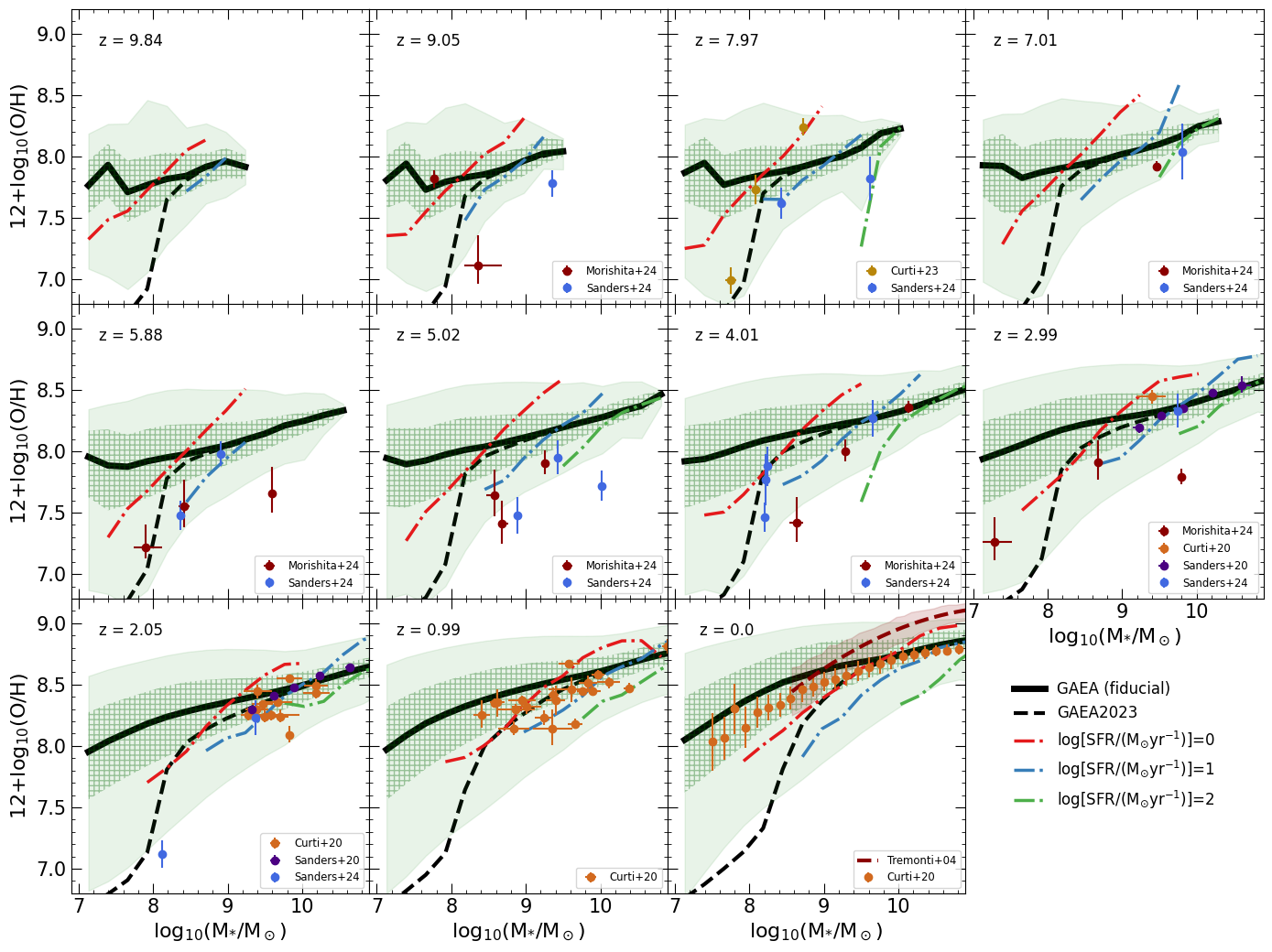}
    \caption{Mass-metallicity Relation (MRZ) up to $z = 10$. GAEA model predictions (lines) are compared with observational data (as indicated in the legends within each sub-panel). The black solid line is the median value of the relation, while the shaded areas show the boundaries between the $16$-th and the $84$-th percentiles (dark green) and the $1$-st and $99$-th percentiles (light green). The predicted MZR has been downshifted by \SI{0.2}{\dex} (see text). The dot-dashed lines are the predicted MZR at fixed SFR. The black dashed lines show predictions from the model published in \citet{delucia2024}, with the only difference being the chemical enrichment in low-mass haloes (see text for details).}
    \label{fig:GAEA_MZR_fixSFR_fiducial}
\end{figure*}

The shape and normalization of the MZR depend significantly on the metallicity tracer and calibration adopted \citep[for a summary, see][]{kewley_ellison2008}. In this work, we compare our predicted MZR with metallicity estimates based on the `direct' method. Also known as the `$T_e$ method', it measures the O/H abundance through the ratio between the [\OIII]$_{\lambda4363}$ auroral emission line and the [\OIII]$_{\lambda5007}$ line. Due to the intrinsic weakness of the [\OIII]$_{\lambda4363}$ line, this metallicity estimate requires high-resolution and high-S/N spectra. The method provides a robust estimate of  the O/H abundance, but depends on the photoionization model adopted for the calculation \citep{gutkin2016}. 

\autoref{fig:GAEA_MZR_fixSFR_fiducial} shows the predicted MZR for $0 \leq z \leq 10$ model galaxies that have a gas fraction $M_{gas} / M_{*} \geq 0.1$ (solid black line). This median MZR from GAEA is downshifted by \SI{0.2}{\dex}. This downshift, that is arbitrary and has been introduced to reproduce the normalization of the observed relation, accounts for the normalization uncertainties discussed above. From dark to light green, the shaded areas represent the  $1 \sigma$ and $3 \sigma$ dispersions of the downshifted median relation, respectively. The black dashed line is the median MZR predicted assuming, as in previous versions of the model, that metals ejected in haloes less massive than $\hat{M}_{vir} = 3 \times 10^{10} \SI{}{\solarmass}$ are mixed with the hot gas (this is also downshifted by \SI{0.2}{\dex}). As can be appreciated from the figure, this modification only affects galaxies with $M_* \lesssim 10^{9}$ \SI{}{\solarmass}, i.e. below the mass resolution of the $N$-body simulation that we used to calibrate the model, so that predictions published in earlier work are hardly affected.  The former prescription underpredicts the metallicity of low-mass galaxies by at least $\sim \SI{1}{\dex}$. In contrast, predictions from the model presented in this study exhibit an approximately log-linear relation, whose normalization increases at decreasing redshift and whose slope is positive and approximately constant in time (see below).

We compare model predictions with different observational estimates based on the direct $T_{e}$ method, and with the MZR measured by \cite{tremonti2004} (dashed red line), based on strong emission lines. Our model predictions are in very good agreement with the local MZR estimated by \cite{curti2020}. When the \SI{0.2}{\dex} downshift is not applied, model predictions are instead in good agreement with estimates  by \cite{tremonti2004}. Model predictions are in agreement with observational estimates up to $z \sim 4$  for galaxies with $\log_\mathrm{10}(M_* / M_\odot) \gtrsim 8$, as discussed in \cite{fontanot2021}. At $5 \leq z \leq 9$, the predicted MZR still exhibits a linear proportionality between the galaxy stellar mass and their gaseous metallicities, matching observational data within the uncertainties. We highlight here that high-$z$ measurements are available only for few  individual systems, while the model MZR shown in Figure \ref{fig:GAEA_MZR_fixSFR_fiducial} is a \textit{median} relation based on a non-negligible (but low) gas fraction.  We have verified that the shape of the predicted MZR does not depend significantly on the adopted gas fraction threshold (see below). 

\begin{figure*}
    \centering
    \includegraphics[width=\linewidth]{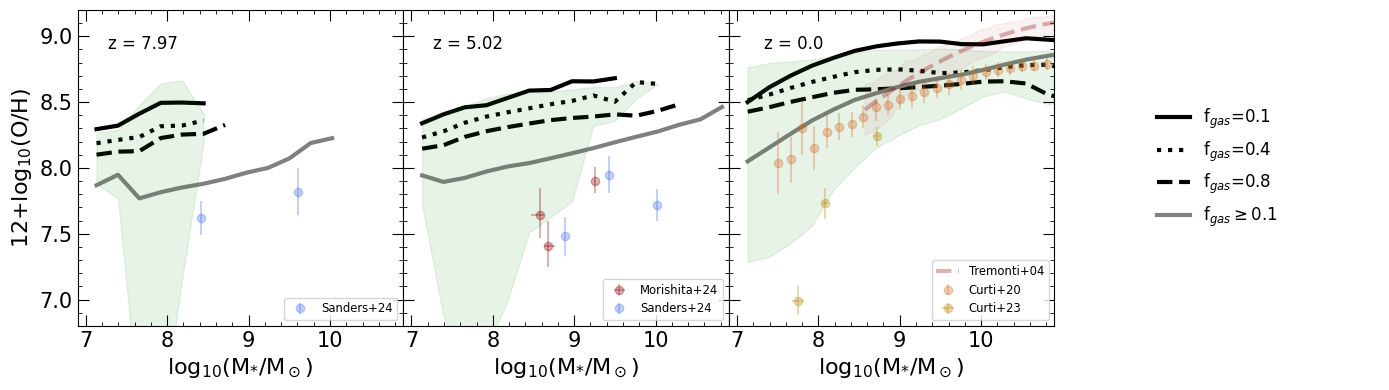}
    \caption{Median MZR at $z = 8, 5, 0$ predicted by GAEA, compared with observational data (see legends). The solid, dotted and dashed lines refer to galaxies having $f_{gas} = 0.1, 0.4, 0.8$, respectively. The grey solid line is the median MZR shown in \autoref{fig:GAEA_MZR_fixSFR_fiducial}. The shaded area corresponds to the $3 \sigma$ contour of the MZR at $f_{gas} = 0.4$. All model predictions have been downshifted by \SI{0.2}{\dex}.}
    \label{fig:GAEA_MZR_fgas_fiducial}
\end{figure*}

We also study the secondary dependencies of the MZR for our model galaxies. First, as discussed in \cite{delucia2020} and \cite{fontanot2021}, we confirm that the MZR at each cosmic epoch can be seen as the superposition of different MZRs at fixed SFR -- namely, the fundamental metallicity relations (FMRs). These are shown in \autoref{fig:GAEA_MZR_fixSFR_fiducial} with dotted-dashed lines. We confirm that the FMR has a negligible evolution  up to $z \sim 3$ and that the evolution of the slope is negligible up to $z \sim 10$ at fixed SFR, with the normalization evolving at $z < 4$.

When selecting only galaxies within a given range of gas fraction -- i.e., $(0.01 - f_{th}) \leq f_{gas} < (f_{th} + 0.01)$, we show in Figure \ref{fig:GAEA_MZR_fgas_fiducial} that the normalization of the median MZR decreases as $f_{gas}$ increases and is above the median MZR computed using all galaxies with $f_{gas} \geq 0.1$. This feature holds at every redshift and up to very large gas fractions. Our results are consistent with \cite{bothwell2013}, who showed that gas-rich galaxies tend to populate the region above the median MZR. However, the scatter is large, and the $3 \sigma$ contours of the MZRs at fixed $f_{gas} = 0.1, 0.4, 0.8$ largely overlap and include model galaxies below the median MZR.

Recent studies have analysed the shape of the MZR in the mass range $10^7$\SI{}{\solarmass}$-10^{10}$\SI{}{\solarmass} at high redshift \citep[see, e.g.,][]{nakajima2023, heintz2023a, morishita2024}. In particular, \cite{morishita2024} considered $25$ galaxies at $3 \leq z \leq 9$ for which the [\OIII]$_{\lambda4363}$ auroral emission line is detected. They find that the observed relation is well fit by the following formula\begin{equation}\label{eq:MZR_fit}
        12 + \log_\mathrm{10}\mathrm{(O/H)} = \alpha \log_\mathrm{10}(M_* / M_0) + B(z) + N(0, \sigma_\mathrm{MZR}),
    \end{equation}
where $\alpha$ is the slope, $B(z) = \alpha_z \log_{10}(1 + z) + \beta_z$ is the redshift-dependent normalization of the MZR and $\sigma_\mathrm{MZR}$ represents the intrinsic scatter of the relation. In the following, we attempt a quantitative comparison of our model predictions with results from \cite{morishita2024}. We start by randomly extracting $10,000$ samples of $25$ model galaxies each, with stellar mass and redshift distributions matching those of the observational sample. We then fit the MZR for each mock catalogue through Equation \ref{eq:MZR_fit}, neglecting the redshift evolution of the normalization term $B(z)$. This choice is motivated by the fact that \citet{morishita2024}: \textit{i)}  found no redshift evolution in their $3 \leq z \leq 9$ sample, with $\alpha_{z,M24} = 0.03^{+0.49}_{-0.51}$; \textit{ii)}  computed the MZR fit considering all the observed galaxies regardless of their redshift. Moreover, the sample is small and splitting it in redshift bins would make the numbers even smaller. When refitting the data in \cite{morishita2024},  we find  $B_{\mathrm M24} (z = 3) = 7.73^{+0.51}_{-0.50}$ and $B_{\mathrm M24} (z = 9) = 7.74^{+0.65}_{-0.65}$, confirming the negligible redshift evolution reported in the original work.

\begin{figure*}[ht]
    \centering
    \begin{subfigure}{0.48\hsize}
        \centering
        \includegraphics[width=\linewidth]{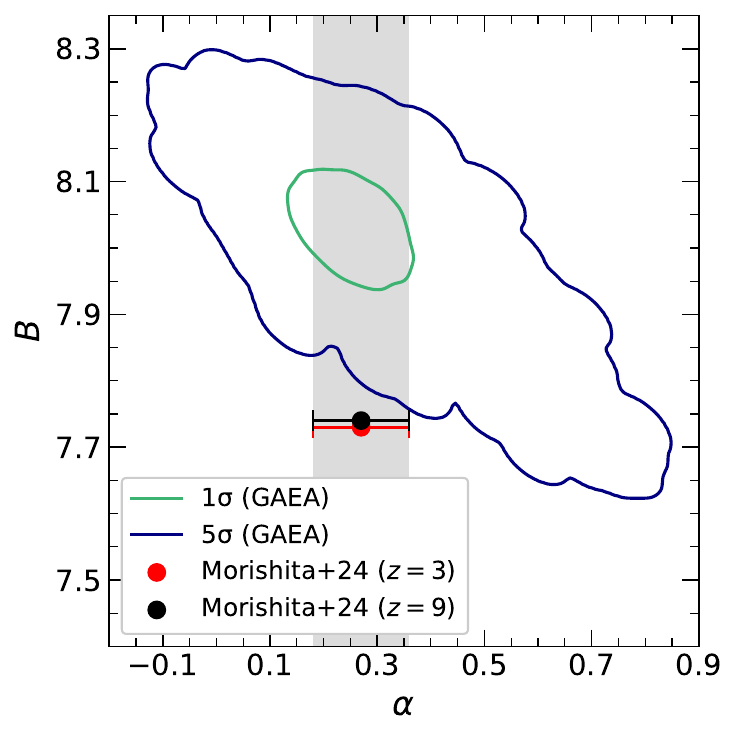}
    \end{subfigure}
    \hfill
    \begin{subfigure}{0.48\hsize}
        \centering
        \includegraphics[width=\linewidth]{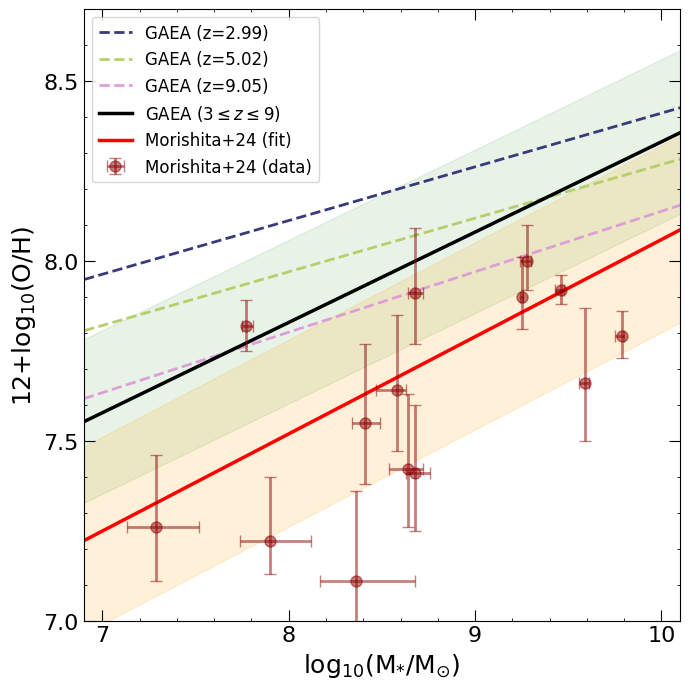}
    \end{subfigure}
    \caption{\textit{(Left)} Distribution of the slope and normalization $(\alpha, B(z))$ for the MZR fit computed using the $10,000$ samples built from the GAEA outputs (see text for details). The $1 \sigma$ (green) and $5 \sigma$ (blue) contours of the distributions are shown. The red and black points with error bars are related to $(\alpha, B(z = 3))$ and $(\alpha, B(z = 9))$ found by \cite{morishita2024}. \textit{(Right)} Fits of the MZR found for GAEA (black solid line), obtained extracting 10,000 mock samples of galaxies that follow the $(M_*, z)$ distribution of the observational data in \cite{morishita2024}. The red solid line is the $3 \leq z \leq 9$ MZR from \cite{morishita2024}, as a comparison. Shaded areas show the scatter of the best fit relations. Dashed lines show the predicted MZR for all the model galaxies with $M_* \geq 10^{8}$ \SI{}{\solarmass} at a fixed redshift -- see legend.}
    \label{fig:MZR_withMorishita_et_al_24}
\end{figure*}

\begin{table*}[hbtp]
  \caption{Parameters of the MZR fit to GAEA galaxies at $z=3, 5, 9$, compared to estimates by \citet{morishita2024} and estimates obtained considering $10,000$ mock catalogues of $25$ model galaxies that match the stellar mass and redshift $(M_*, z)$ distribution sample of \cite{morishita2024}. For the model fit, a downshift of \SI{0.2}{\dex} in metallicity is considered.}
  \label{tab:MZR_fit_params}

  \renewcommand{\footnoterule}{}
  \renewcommand{\arraystretch}{1.2}
  \centering
  \begin{tabular}{ccccccc}
\hline\hline
\multicolumn{6}{l}{Fit from \cite{morishita2024}.} \\
\hline
$z$ & $\log_{10}(M_0 / M_\odot)$ & $\alpha$    & $\beta_z$ & $\alpha_z$ & $\sigma_{\log \textup{MZR}}$ \\ 
$3 \leq z \leq 9$ & $8.8$ & $0.27^{+0.09}_{-0.09}$ & $7.71^{+0.42}_{-0.40}$ & $0.03^{+0.49}_{-0.51}$ & $0.26^{+0.06}_{-0.04}$\\ \hline\hline
\multicolumn{6}{l}{Fit from $10,000$ mock catalogues of $25$ model galaxies from GAEA.} \\
\hline
$z$ & & $\alpha$ & $B(z) = \beta_z + \alpha_z(1 + z)$ & & $\sigma_{\log \textup{MZR}}$ \\
$3 \leq z \leq 9$ & & $0.251 \pm 0.001$ & $8.0296 \pm 0.0004$ & & $0.2274 \pm 0.0003$\\ 
\hline\hline
\multicolumn{6}{l}{Fit from GAEA's model galaxies with $\log_{10}(M_* / M_\odot) \geq 8$.} \\ \hline
$z$ & & $\alpha$ & $B(z) = \beta_z + \alpha_z(1 + z)$ & & $\sigma_{\log \textup{MZR}}$\\ 
3 & & $0.15$ & $8.23$ & & 0.26\\
5 & & $0.15$ & $8.09$ & & 0.26\\
9 & & $0.17$ & $7.93$ & & 0.23\\
\hline\hline
  \end{tabular}
\end{table*}

The distribution of slopes and normalizations for our mock catalogues  is shown in \autoref{fig:MZR_withMorishita_et_al_24} (left). The green and blue contours represent the $1\sigma$ and $5\sigma$ contours of the distribution, respectively. The red and black points with the error bars correspond to the values $(\alpha_{\mathrm M24},B_{\mathrm M24}(z=3))$ and $(\alpha_{\mathrm M24},B_{\mathrm M24}(z=9))$ that we have computed  refitting the observational estimates. About $74\%$ of our mock catalogues have a slope consistent with that found by \citeauthor{morishita2024}  ($\alpha_\mathrm{M24}=0.27\pm0.09$) and, due to the large uncertainties in $B_\mathrm{M24}(z)$, $100\%$ of our mock samples have a normalization comparable to their $B_\mathrm{M24}(z)$ at $3 \leq z \leq 9$, although the mock average central values is \SI{0.3}{\dex} higher than the estimated one. The distribution of $(\alpha,B(z))$ also matches the observational findings, with a $89\%$ probability of having $(\alpha_\mathrm{M24},B_\mathrm{M24}(z))$ within the uncertainty range. In addition, when the average slope and intercept are compared from the $(\alpha,B(z))$ distributions obtained for the mock catalogues from GAEA, they are in good agreement with the estimates by \cite{morishita2024}: $\alpha_\mathrm{GAEA}=0.251\pm0.001$ and $B_\mathrm{GAEA}(3\leq z\leq 9)=8.0296\pm0.0004$. 

\autoref{fig:MZR_withMorishita_et_al_24} (right) shows the predicted fit for the MZR and that obtained by \cite{morishita2024}. 
The solid black line represents the average fit of the predicted MZR from GAEA using the $10000$ mock catalogues. The red points with error bars and the red solid line represent the observational data and the fit from \cite{morishita2024}, respectively. Shaded areas show the scatter of the relations. The predicted and observed relations agree well both in terms of slope and scatter values (we refer to \autoref{tab:MZR_fit_params} for further details), but not for the normalization: model galaxies have metallicities roughly \SI{0.3}{\dex} larger than observational estimates.

Linear fits of the MZR at fixed redshift ($z=3,5,9$) using all model galaxies with $M_*\geq10^{8}$ \SI{}{\solarmass} are also shown in \autoref{fig:MZR_withMorishita_et_al_24} (right). We clearly see a redshift evolution of the normalization: from $z=9$ to $z=3$ it increases by approximately $\SI{0.3}{\dex}$. On the other hand, the slope and the intrinsic scatter of the relation do not significantly evolve in this redshift range -- see \autoref{tab:MZR_fit_params}. When the intrinsic scatter of the fit is combined with its normalization, the redshift evolution becomes weaker. This conclusion is shared by recent independent observational works (e.g., \cite{matthee2023} at $z \sim 6$; \cite{heintz2023a} at $7 \leq z \leq 10$; \cite{curti2023} at $3 \leq z \leq 10$; \cite{morishita2024} at $3 \leq z \leq 9$).

\section{Are the brightest and most massive galaxies in the very-high redshift Universe breaking $\Lambda$CDM?}\label{sec:model_variants}

\begin{figure*}
	\centering
	\includegraphics[width=\linewidth]{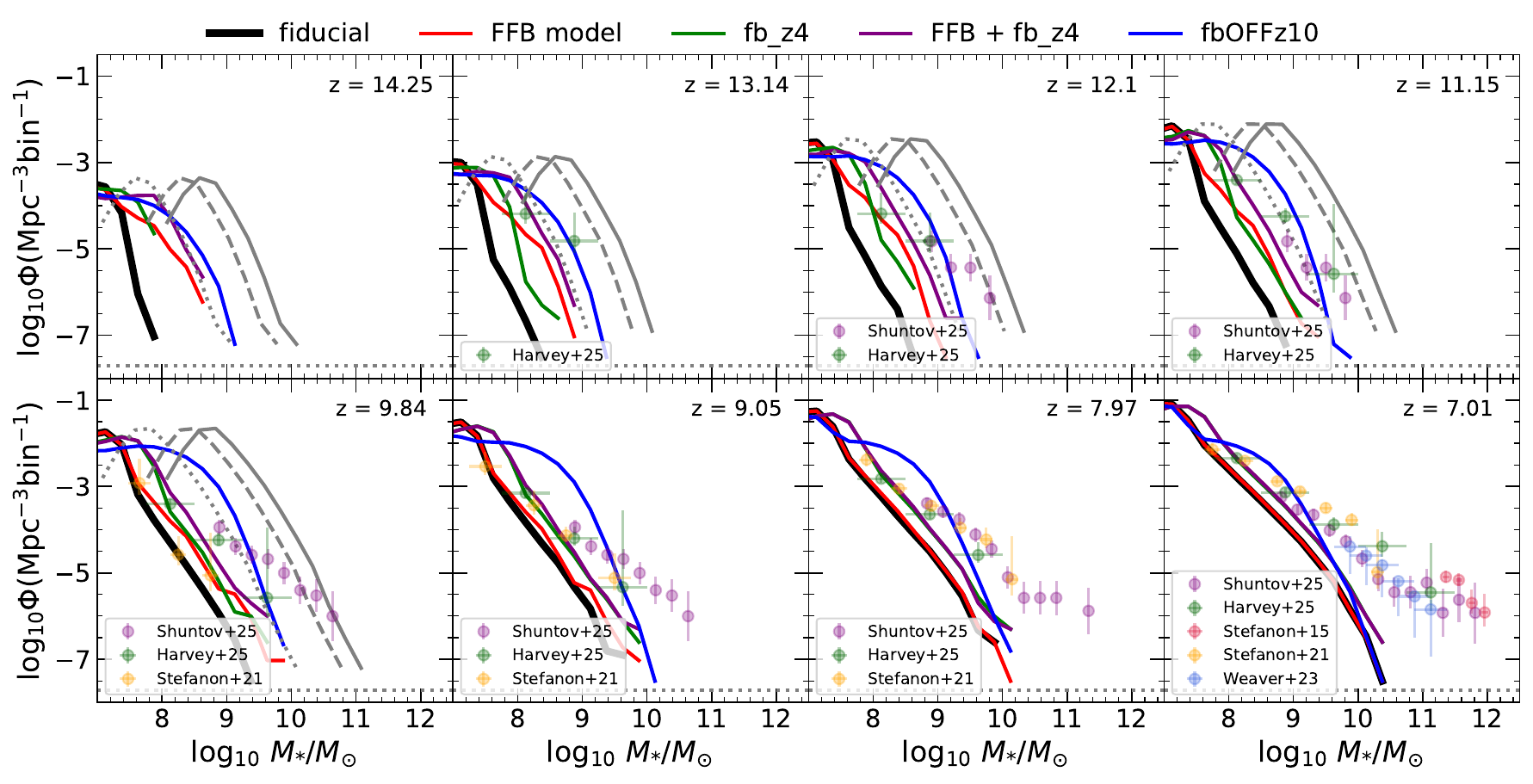}
	\caption{Same legend as in \autoref{fig:GAEA_GSMF_fiducial}, but including predictions from model variants: \textit{a)} FFB model at $z > 10$ (red line); \textit{b)} redshift-independent stellar feedback at $z > 4$ (green line); \textit{c)} combination of \textit{a)} and \textit{b)} (purple line); \textit{d)} suppressed stellar feedback  at $z > 10$ (blue line). Grey lines at $z \gtrsim 10$ show predictions from a  HOD model assuming star formation  efficiencies $\epsilon = 10\%$ (dotted), $\epsilon = 50\%$ (dashed) and $\epsilon = 100\%$ (solid).}
	\label{fig:GAEA_GSMF_variants}
\end{figure*}

\begin{figure*}
	\centering
	\includegraphics[width=\linewidth]{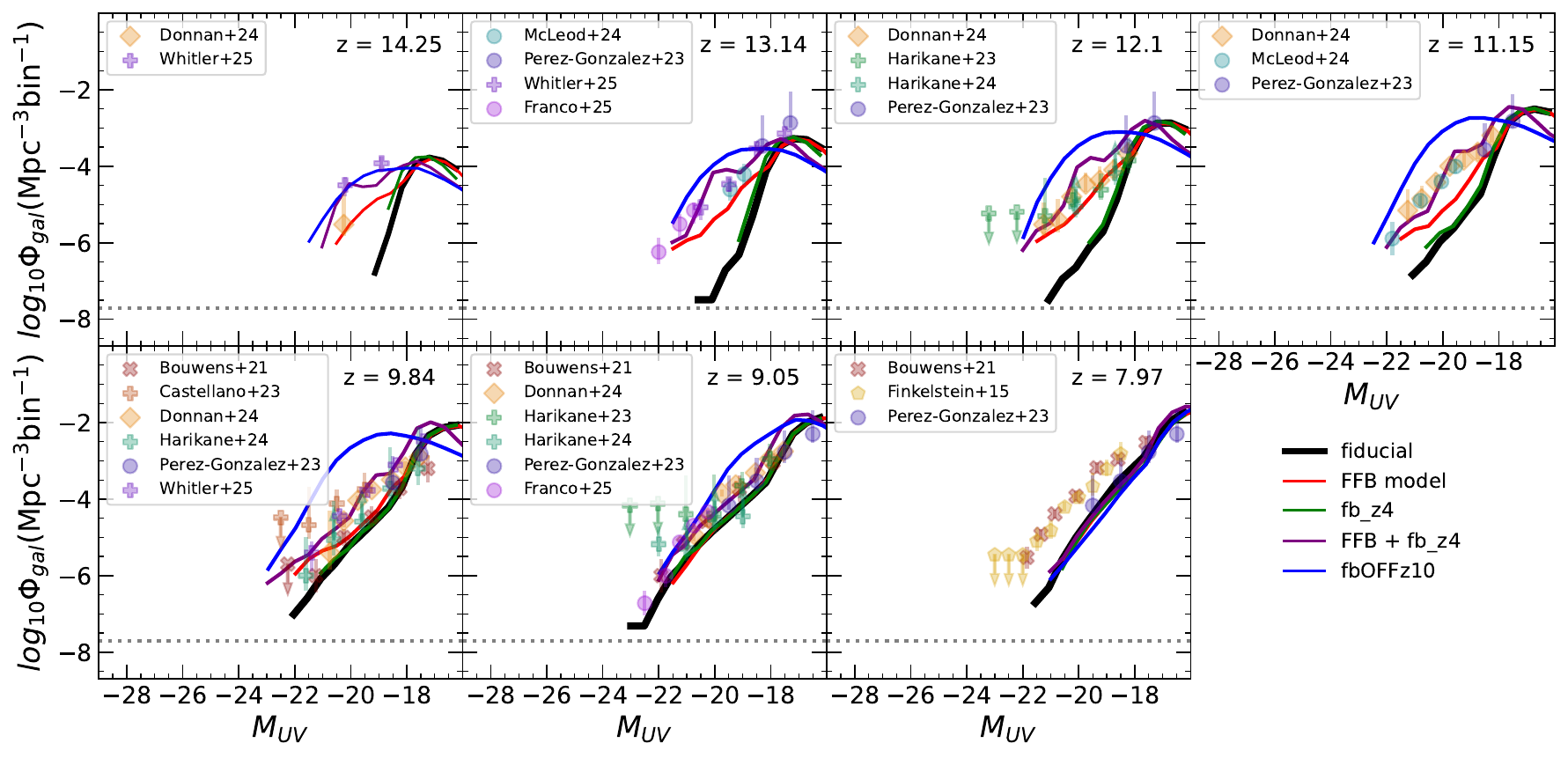}
	\caption{Same legend as in \autoref{fig:GAEA_galUVLF_fiducial}, but including model variants. The color code is as in \autoref{fig:GAEA_GSMF_variants}.}
	\label{fig:GAEA_galUVLF_variants}
\end{figure*}

As seen in Sections \ref{sec:GAEA_fiducial_GSMF} and \ref{sec:GAEA_fiducial_UVLF}, model predictions underestimate the number density of bright and UV-luminous galaxies at $z > 10$ with respect to observational data. We address this discrepancy by considering different physical explanations: \textit{i)} inefficiency of star formation; \textit{ii)} too strong stellar feedback; \textit{iii)} a combination of \textit{i)} and \textit{ii)}. 

We have also considered  the impact of a different redshift of reionization, but found that this is not important for the high-mass end of the GSMF nor for the bright end of the UVLF. In fact, our fiducial model assumes an `early' reionization starting at $z=15$ and ending at $z = 11.5, $with a duration of \SI{122}{\mega\year} \citep{hirschmann2016}. This scenario was broadly consistent with earlier constraints from the WMAP satellite \cite[e.g.,][]{wmap2013}. However, more recent measurements from the Planck satellite have revised this value downwards, favouring a later and more extended Epoch of Reionization \cite[e.g.,][]{planck2018}. To test the impact of this shift in the reionization epoch, we explore a model variant with a reionization starting at $z = 10$ and ending at $z = 6$ (with a duration of \SI{460}{\mega\year}). Our findings indicate that varying the redshift range for reionization primarily enhances the number density of low-mass galaxies ($M_* < 10^8$ \SI{}{\solarmass})  at $3 \leq z \leq 9$ and of UV-faint galaxies ($M_{UV} \gtrsim -18$) at $6 \leq z \leq 9$. The enhancement is small (of the order of \SI{0.5}{\dex}) so that model predictions in these mass and luminosity ranges remain  in reasonable agreement with observational data. 

\subsection{Star formation efficiency at $z > 10$}\label{sec:FFB_model}
In our reference run, we assume that the surface density of star formation $\Sigma_{\textup{sf}}$ is proportional to the surface density of the molecular gas $\Sigma_{\textup{H}_2}$ through
\begin{equation}
    \label{eq:star_formation_law}
    \Sigma_{\textup{sf}} = \nu_{\textup{sf}} \Sigma_{\textup{H}_2},
\end{equation} 
where $\nu_{\textup{sf}}$ is the star formation efficiency timescale (in units of \SI{}{\per\giga\year}). For details about this implementation, we refer to \cite{xie2017}. 

This model variant represents an attempt to implement the `feedback-free starburst' (FFB) model that has been proposed by \cite{dekel2023} and \cite{li2024}. The authors argue that, at $z \sim 10$ and above, high-density and metal poor gas leads to an elevated star formation efficiency in the central regions of massive halos (e.g., for $M_h \gtrsim 10^{11}$\SI{}{\solarmass}). For these conditions of density and metallicity, the free fall time of the gas becomes shorter than $\sim$\SI{1}{\mega\year}, which is the timescale required for low-metallicity massive stars to release significant amounts of energy into the ISM through winds and supernovae. This timescale can be translated into a characteristic density of $\tilde{n} = 3 \times 10^3$\SI{}{\per\cm\cubed} above which FFBs occur. In our implementation of the FFB model, we adopt $\tilde{n}$ as a threshold above which in all $z \gtrsim 10$ galaxies the star formation timescale is set to \SI{1}{\mega\year}. Moreover, we assume that the total mass of newly formed stars and the reheated mass cannot be larger than the total cold gas mass inside the halo. This assumption sets a limit to the star formation efficiency.

Results from this model version are shown in \autoref{fig:GAEA_GSMF_variants} and in \autoref{fig:GAEA_galUVLF_variants}, together with predictions from other model variants discussed in the following subsections. The red lines represent the GSMF obtained by switching on the FFB model at $z > 10$: this assumption increases significantly the number density of massive galaxies beyond $z\sim10$, bringing model predictions in a better agreement with observational estimates. At $z < 10$, the predicted GSMF converges to the one predicted by our fiducial model. As for the UVLF, results shown in \autoref{fig:GAEA_galUVLF_variants} are in very good agreement with observational data: also in this case, the number densities of UV-bright galaxies are larger than those based on the fiducial model, by about \SI{2}{\dex}. 

We stress that the FFB model is not only physically motivated, but also returns a realistic star formation efficiency. To show that, we compare the GSMF given by this model variant to HOD stellar mass functions, shown in \autoref{fig:GAEA_GSMF_variants} as grey lines. To compute them, we consider all the gas inside the virial radius of the halo, $M_h f_b$, and multiply it by a constant star formation efficiency $\epsilon$. In \autoref{fig:GAEA_GSMF_variants}, we show the resulting lines for $\epsilon = 10\%$ (dotted), $\epsilon = 50\%$ (dashed) and $\epsilon = 100\%$ (solid). The FFB model enhances the global star formation efficiency to $10\%$, which is not an extreme value.

\subsection{Weaker stellar feedback at $z > 4$}\label{sec:weaker_fbz4}
In our fiducial model we parametrize the stellar feedback using the same prescriptions adopted in \cite{hirschmann2016} (Equations $14$ and $15$), based partially on fit to hydro-dynamical simulation results published in \cite{muratov2015}. Specifically, the reheated gas mass rate $\dot{M}_{\textup{reheat}}$ and the energy injection rate by massive stars $\dot{E}_{FB}$ are parametrized as:
\begin{equation}
	\label{eq:reheated_gas_mass_rate}
	\dot{M}_{\textup{reheat}} = \epsilon_{\textup{reheat}}(1 + z)^{1.25} \left( \frac{V_{\textup{max}}}{\SI{60}{\kilo\meter\per\second}} \right)^\alpha \times \dot{M}_{\textup{star}}
\end{equation}
and
\begin{equation}
	\label{eq:energy_injection_rate}
	\dot{E}_{\textup{FB}} = \epsilon_{\textup{eject}}(1 + z)^{1.25} \left(\frac{V_{\textup{max}}}{\SI{60}{\kilo\meter\per\second}} \right)^\alpha \times 0.5 \dot{M}_{\textup{star}} V^2_{\textup{SN}},
\end{equation}
respectively. $\epsilon_{\textup{reheat}}$ is the reheating efficiency, $\epsilon_{\textup{eject}}$ is the ejection efficiency, $V_{\textup{max}}$ is the maximum value of the circular velocity of the gas disc, $\dot{M}_{\textup{star}}$ is the star formation rate, and $0.5 \dot{M}_{\textup{star}} V^2_{\textup{SN}}$ is the mean kinetic energy of the SN ejecta per unit mass of stars formed. The parametrizations found by \cite{muratov2015} are derived from numerical experiments valid up to $z = 4$. Our reference run extrapolates the trend to higher redshifts. This implies that the stellar feedback increases with redshift at fixed values of $\epsilon_{\textup{reheat}}$, $V_{\textup{max}}$ and $\dot{M}_{\textup{star}}$. In this model variant, we assume that \autoref{eq:reheated_gas_mass_rate} and \autoref{eq:energy_injection_rate} hold up to $z = 4$, and the values of $\dot{M}_{\textup{reheat}}$ and $\dot{E}_{FB}$ are then kept constant at higher redshifts at fixed $\epsilon_{\textup{reheat}}$, $V_{\textup{max}}$ and $\dot{M}_{\textup{star}}$.

Green lines in \autoref{fig:GAEA_GSMF_variants} and in \autoref{fig:GAEA_galUVLF_variants} show the effect of this modification. We find that a weaker stellar feedback increases the normalization of the $z \gtrsim 8$ GSMF. However, this increase in the number densities of the most massive galaxies does not translate into an increase in the number densities of the brightest galaxies. So in this model variant, more stars are formed at high redshift but these do not lead to an increased UV luminosity. This can be explained by considering the star formation law in Equation \ref{eq:star_formation_law} and its related time scale in Figure~$3$ from \cite{xie2017}: the conversion time scale to form new stars from molecular hydrogen is of the order of $10^{8-9}$ \SI{}{\giga\year} even for gas surface densities of $10^{3-4}$ \SI{}{\solarmass\per\parsec}. Such time scale is two orders of magnitude larger than the average lifetime of low-metallicity massive stars (a few million years -- these are the main contributors to the $z \gtrsim 10$ UVLF). We conclude that, in the framework of our model, a weaker stellar feedback is not sufficient alone to reproduce the observed number density of high-$z$ bright galaxies.

\subsection{Weaker stellar feedback and more efficient star formation}\label{sec:comb_FFB_fbz4_variant}
We now investigate the combined effect of a weaker stellar feedback at $z > 4$ and the FFB model, that we assume holds at $z \gtrsim 10$. Results are shown by the purple lines in \autoref{fig:GAEA_GSMF_variants} and in \autoref{fig:GAEA_galUVLF_variants}. This model variant yields larger number densities for both the GSMF and the UVLF. In particular, the enhanced star formation efficiency at $z \gtrsim 10$ and redshift-saturated feedback at $z > 4$ returns a slightly larger number density of massive and bright galaxies than when these variants are considered independently. As expected from the previous discussion, the main contribution in this combination is given by the FFB model, that boosts the star formation and UV luminosity emission. The model results based on this model variant exhibit an improved agreement with current observational estimates, and the related GSMF overlaps the HOD SMF with a star formation efficiency of about ten percent.

\subsection{Suppressing stellar feedback at $z > 10$}\label{sec:no_feedback_z10_model}
As a final test, we simply shut down gas reheating and the energy injection  at $z > 10$. At $z \gtrsim 10$, the FFB scenario predicts extremely dense, compact star-forming clouds whose free-fall times are shorter than the timescales for massive-star winds and supernovae to develop. Therefore, radiative and mechanical reheating by stars is expected to be largely ineffective.

Predictions from this model variant are shown in \autoref{fig:GAEA_GSMF_variants} and in \autoref{fig:GAEA_galUVLF_variants} with blue lines. The resulting GSMF and UVLF exhibit a significantly larger number of massive and bright galaxies beyond $z\sim9$ with respect to our fiducial model, slightly larger than those coming from the combined model variant described in Section \ref{sec:comb_FFB_fbz4_variant}. The boosted production of bright and massive galaxies is equivalent to that of a HOD SMF with a star formation efficiency slightly larger than ten percent, though it is degenerate with the combined variant described in Section \ref{sec:comb_FFB_fbz4_variant}. Predictions from this model variant are in very good agreement with $z>10$ data, while they exceed the observed number density below that redshift, especially at the low-mass end, and converge with the fiducial results at $z \sim 7$. This model variant predicts a UVLF that is in good agreement with observations at every redshift, and converges with the fiducial UVLF at $z = 8$. 

\subsection{Can we discriminate between different physical modifications?}

\begin{figure*}
	\centering
	\includegraphics[width=0.95\linewidth]{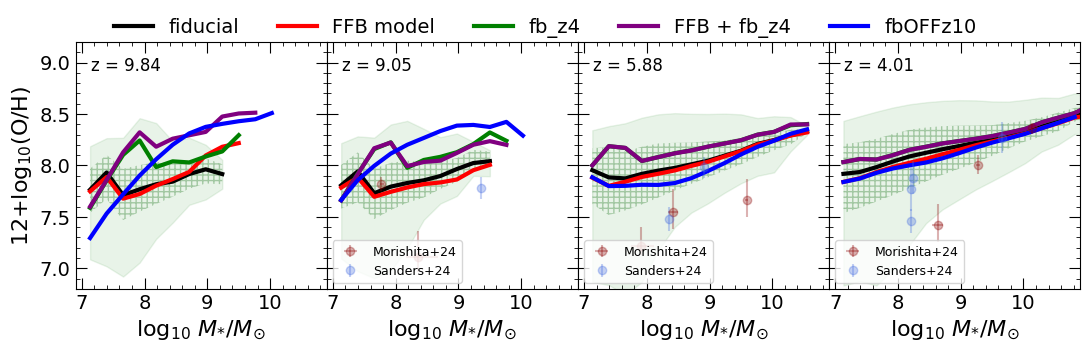}
	\caption{Same legend as in \autoref{fig:GAEA_MZR_fixSFR_fiducial}. The color code for the lines is the same as in \autoref{fig:GAEA_GSMF_variants}.}
	\label{fig:GAEA_MZR_variants}
\end{figure*}

\begin{figure*}
	\centering
	\includegraphics[width=0.95\linewidth]{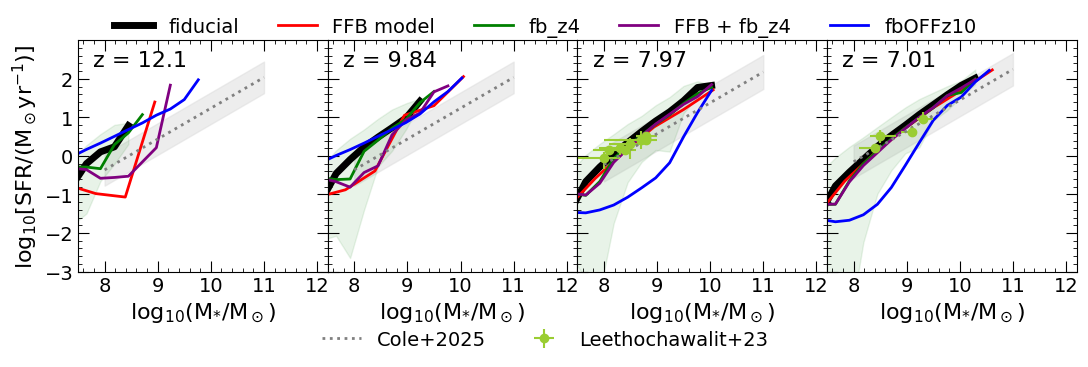}
	\caption{SFR-stellar mass relation predicted by GAEA. The lines are the median relation for each variant, while the green shaded areas are $3 \sigma$ contours. The color code for solid lines is the same as in \autoref{fig:GAEA_GSMF_variants}. For comparison, observational data (green circles) from \cite{leethochawalit2023} and fit (grey dotted line with shaded areas) from \cite{cole2025} are shown.}
	\label{fig:GAEA_SFR_Mstar_variants}
\end{figure*}

Predictions from the `FFB+$z$-saturated feedback' model and from the `no high-$z$ stellar feedback' model (Sections \ref{sec:comb_FFB_fbz4_variant} and \ref{sec:no_feedback_z10_model}, respectively) are degenerate in terms of mass function and UV luminosity function, in particular when considering observational uncertainties. However, the degeneracy can be broken by considering the impact of the different physical modifications on additional physical properties.

A straightforward consequence of enhancing the formation of new stars is an increase of the metal production and ejection in the ISM, although this could be counter-balanced by a dilution of the global metallicity due to the infall of pristine gas. \autoref{fig:GAEA_MZR_variants} shows model predictions for the variants discussed in Section \ref{sec:GAEA_fiducial_MZR}. All model variants predict larger gaseous metallicities at $z > 5$. In particular, galaxies from the `no high-$z$ stellar feedback' model have the highest metallicities at $8 < z < 9$, while at $5 < z < 7$ galaxies from the `FFB+$z$-saturated feedback' model are more metal-rich than those of all the other models by a factor of \SIrange{0.1}{0.2}{\dex}. Thus, all model variants considered reinforce the offset with respect to high-$z$ observational estimates discussed above. It is useful to note that,  when the scatter in the MZR and observational uncertainties are taken into account, the differences between the model variants considered become less significant. 

The enhancement of the star formation, changing either the star formation time scale or the prescription for stellar feedback, can also lead to different behaviours in star formation histories. \autoref{fig:GAEA_SFR_Mstar_variants} shows the relation between the star formation rate and the galaxy stellar mass (equivalent trends are visible when considering the relation between cold gas mass and galaxy stellar mass). Beyond $z \sim 7$ we can distinguish the behaviour of two variants -- the `no high-$z$ stellar feedback' model and the FFB model. In particular, galaxies with $\log_{10}(M_* / M_\odot) \lesssim 9$ in the former model have less gas than galaxies in the fiducial model (with a gas mass content that is lower than the fiducial relation by of \SIrange{1}{1.5}{\dex}) at $7\lesssim z \lesssim 9$, while the FFB model shows the same behaviour at $z > 10$. These different behaviours can be explained as follows: first, for the FFB model, the cold gas content inside galaxies is low because it is rapidly converted into stars on a \SI{}{\mega\year}-timescale; second, in the `no high-$z$ stellar feedback' model the amount of cold gas inside galaxies is larger at early epochs, leading to higher star formation and earlier black hole accretion, quickly depleting the gas mass reservoir at more recent cosmic epochs.

Model predictions are compared with observational data from \cite{leethochawalit2023} and fit from \cite{cole2025}. In particular, from the latter work, we show the SFR$_{10} - M_*$ fit, where SFR$_{10}$ is the star formation history averaged on a time scale of \SI{10}{\mega\year}. Such narrow time interval is comparable with the time separation between two snapshots in the PMS at $z > 7$, that is of the order of \SIrange{10}{20}{\mega\year}. At $z \leq 8$, observations are in agreement with the fiducial predictions, and the median relation from the `no high-$z$ stellar feedback' model is outside the $1 \sigma$ uncertainty of the fit for low mass galaxies, where this model variant predicts lower SFRs than the fiducial ones. Thus, shutting down the stellar feedback in the very-high redshift, while enhancing the $z \geq 10$ stellar mass and UV luminosity functions, results in lower SFRs at lower redshifts, which in principle is not compatible with current observational estimates. However, observational measurements at these early epochs are affected by large uncertainties and potentially important biases. The UV-selected sample from \cite{leethochawalit2023} is biased towards bright and dust-poor galaxies and is incomplete at low stellar masses and low SFRs. Their SFRs correspond to  $\sim$\SI{100}{\mega\year} timescales and may underestimate obscured or bursty star formation.  \cite{cole2025} inferred SFRs from SED fitting, which leads to rather large uncertainties.

Other observed relations can be examined to discriminate between these variations of the model at high redshift. We limit the discussion to the predicted black hole-stellar mass relation in Appendix \ref{app:BH_stellar_mass_relation}.

\section{Summary and conclusions}\label{sec:summary_conclusion}
This work presents a theoretical investigation of the high-redshift Universe using the GAEA state-of-the-art galaxy formation model, run on merger trees from a large cosmological $N$-body simulation. The predicted galaxy stellar mass function is in good agreement with observational estimates at high redshift when observational mass uncertainties of the order of \SIrange{0.25}{0.5}{\dex} are considered. The corresponding stellar-to-halo mass relation agrees well with independent theoretical and empirical models, except for an underprediction of the stellar content in high-mass halos at $z > 2$. Our reference model successfully reproduces the observed UV luminosity function, when considering contributions from both galaxies and AGN, up to $z \simeq 10$. We find that AGN provide a significant contribution to the bright end ($M_{UV} \lesssim -22$) up to $z \sim 8 - 9$. Our dust model also implies a decreasing contribution of dust attenuation at $z \gtrsim 6$. This is consistent with expectations of inefficient dust growth at early cosmic epochs but in contrast with recent  observations  revealing substantial dust attenuation of the cosmic star formation density up to $z \sim 15$. The model also reproduces the mass-metallicity relation up to $z \sim 4$, confirming its known secondary dependencies with SFR and gas fraction. The predicted MZR exhibits a weak evolution in terms of slope, scatter and normalization. The latter is offset by \SI{0.3}{\dex} with respect to recent observational estimates for galaxies at $3 < z < 9$.

We have investigated different physical explanations for the deficit of massive and UV bright galaxies at high redshift predicted by the fiducial GAEA model. In particular, we identify two model variants leading to an increased star formation at high redshift and model predictions that are in much better agreement with observational estimates. The first is a `feedback-free starburst' (FFB) model, in which the star formation efficiency is enhanced at $z>10$ by shortening the star formation timescale in massive halos and preventing the disruption of cold gas clouds inside them. The second is a `no high-$z$ stellar feedback' model, in which stellar feedback at $z > 10$ is suppressed. Both these scenarios are physically motivated and successfully reconcile the predicted GSMF and UVLF with observational estimates. We also discuss how these two scenarios can possibly be discriminated by examining additional key scaling relations. Although both models are degenerate in terms of their predictions for the MZR at $z > 10$, the `no high-$z$ stellar feedback' model predicts a significantly higher metallicity at $z=9$. The SFR-stellar mass relation offers another diagnostic: the `no high-$z$ stellar feedback' model  predicts a lower average SFR at $z < 9$ than our reference model, and is in disagreement with current observational estimates. However, the latter are still affected by large uncertainties and potentially important biases to effectively discriminate among the proposed scenarios. 

We note that there are additional important physical modifications that we have not explored in this work. Our analysis is conducted within the standard $\Lambda$CDM cosmological framework and does not consider alternatives such as early dark energy models \citep{liu2024} that could also lead to an increase in the number densities of bright and massive galaxies at high-$z$. We also assume a universal stellar IMF, whereas an evolving IMF that is more top-heavy at high redshift or for stars forming in metal-poor gas regions could alleviate the observed tensions \citep[see, e.g.,][]{mauerhofer2025}. Finally, the Planck-Millennium simulation that we use for this work lacks the mass resolution to model the formation of Population III stars in the so-called `mini-halos' ($10^5 M_\odot < M_h < 10^6 M_\odot$) at $z>10$, which could contribute significantly to the galaxy population at these early cosmic epochs. We will address these additional elements in future work.

\begin{acknowledgements}
      Information about the model, including details, recent work, and published data can be found at \href{https://sites.google.com/inaf.it/gaea/}{https://sites.google.com/inaf.it/gaea/}. We acknowledge the use of INAF-OATs computational resources within the framework of the CHIPP project \citep{taffoni2020}. MH and AD acknowledge funding from the Swiss National Science Foundation (SNF) via a PRIMA Grant PR$00$P$2$ $193577$ `From cosmic dawn to high noon: the role of black holes for young galaxies’. This project has received funding from the European Union’s Horizon $2020$ research and innovation programme under the Marie Skłodowska-Curie grant agreement No $101148925$.
\end{acknowledgements}

\bibliographystyle{aa}
\bibliography{highz_gaea} 

\begin{appendix} 
\section{Convergence of model predictions as a function of the time-sampling adopted for the merger trees construction}
\label{app:compare_PMS_runs}

\begin{figure*}[h!]
	\centering
	\includegraphics[width=\linewidth]{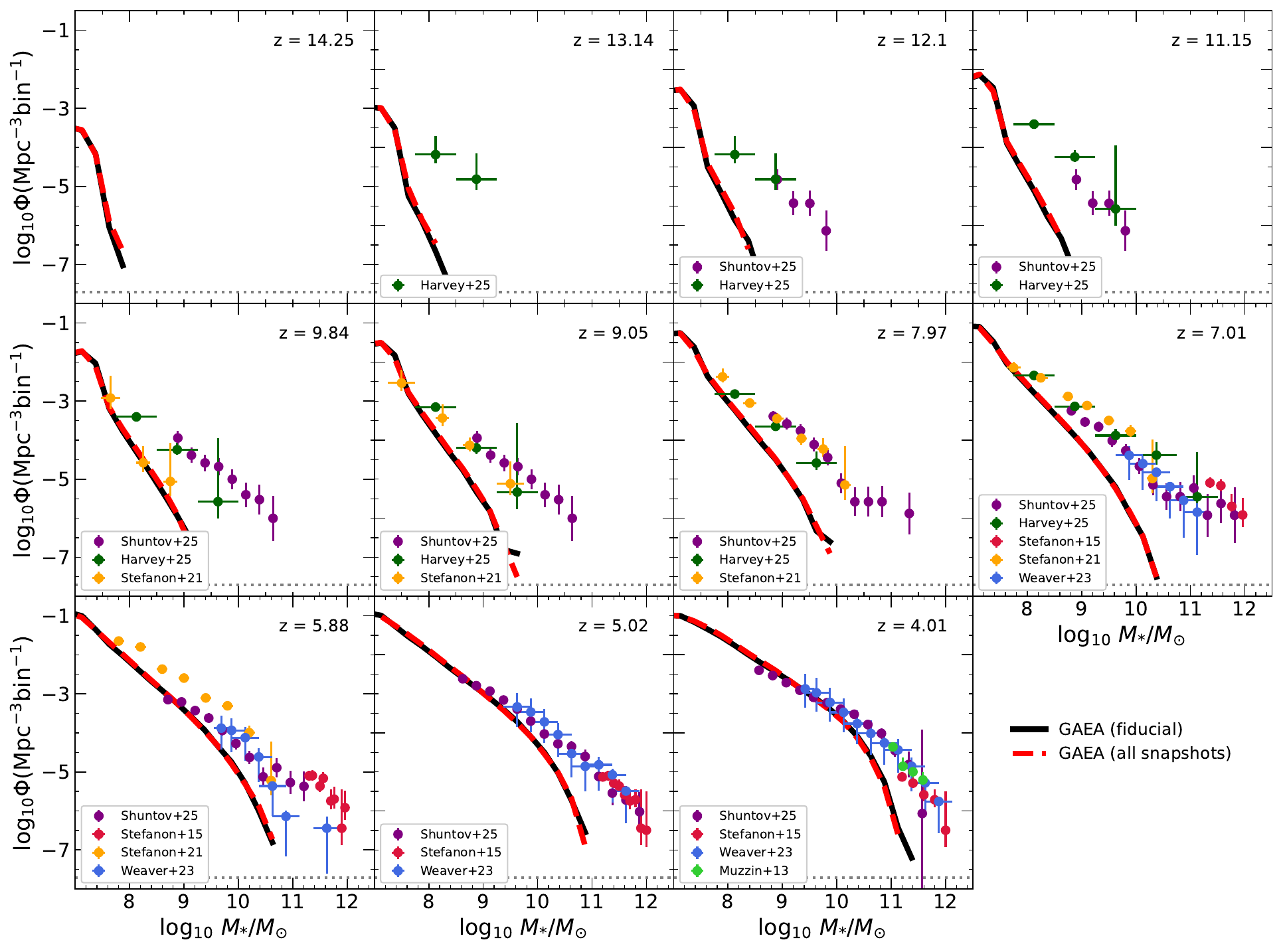}
	\caption{Predicted GSMFs at $14 \leq z \leq 4$ as predicted from GAEA run on two different sets merger trees extracted from the PMS. Solid black lines (fiducial results) are obtained from merger trees constructed using half of the available outputs from the PM; dashed red lines are the GSMFs derived using all available outputs from the PMS. Observational data points are the same as those presented in \autoref{fig:GAEA_GSMF_fiducial}.}
	\label{fig:compare_PMS_runs_GSMF}
\end{figure*}

\begin{figure*}[h!]
	\centering
	\includegraphics[width=\linewidth]{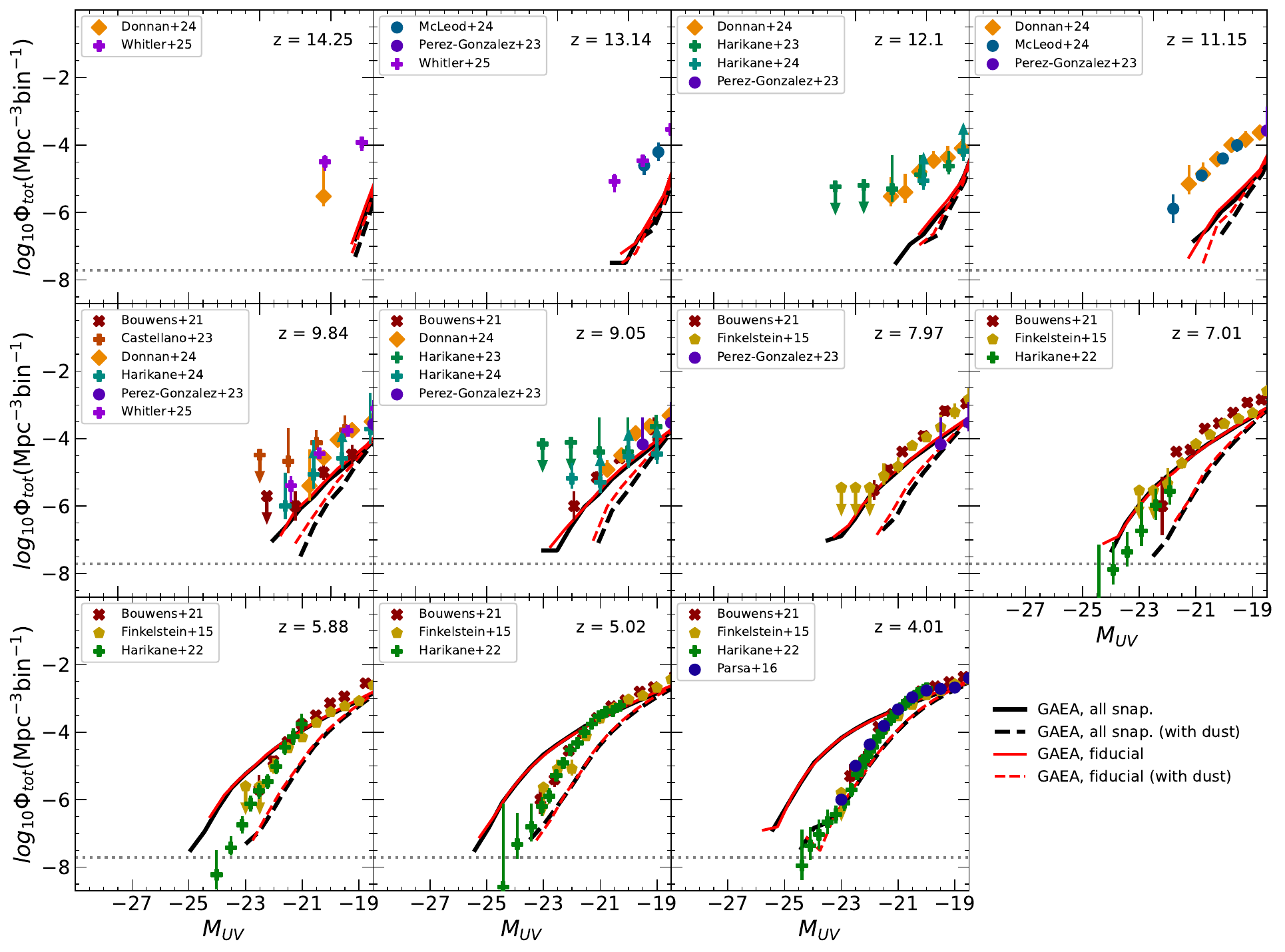}
	\caption{Predicted galaxy UVLFs at $14 \leq z \leq 4$ as predicted from GAEA run on two different sets of merger trees extracted from the PM. Solid black lines (fiducial results) are obtained from merger trees constructed using half of the available outputs from the PM; solid red lines are derived using all available outputs. Dashed black and red lines show the dust-attenuated UVLFs. Observational data points are the same as those presented in \autoref{fig:GAEA_galUVLF_fiducial}.}
	\label{fig:compare_PMS_runs_galUVLF}
\end{figure*}

Figures \ref{fig:compare_PMS_runs_GSMF} and \ref{fig:compare_PMS_runs_galUVLF} show the predicted GSMF and the UVLF at $14 \leq z \leq 4$, based on two different sets of merger trees extracted from the PMS. Black lines are related to the fiducial model, namely the model predictions described in Sections \ref{sec:GAEA_fiducial_GSMF} and \ref{sec:GAEA_fiducial_UVLF}. For this run, merger trees are sampled using one every second available snapshot of the simulation (i.e., half of all available outputs). Red lines show model predictions obtained using merger trees sampled using instead all available outputs of the PMS. The convergence between the two runs is excellent. 

\section{Comparing techniques to compute the AGN UV luminosities}
\label{app:comp_AGN_UVLF_techniques}
\begin{figure*}
	\centering
	\includegraphics[width=\linewidth]{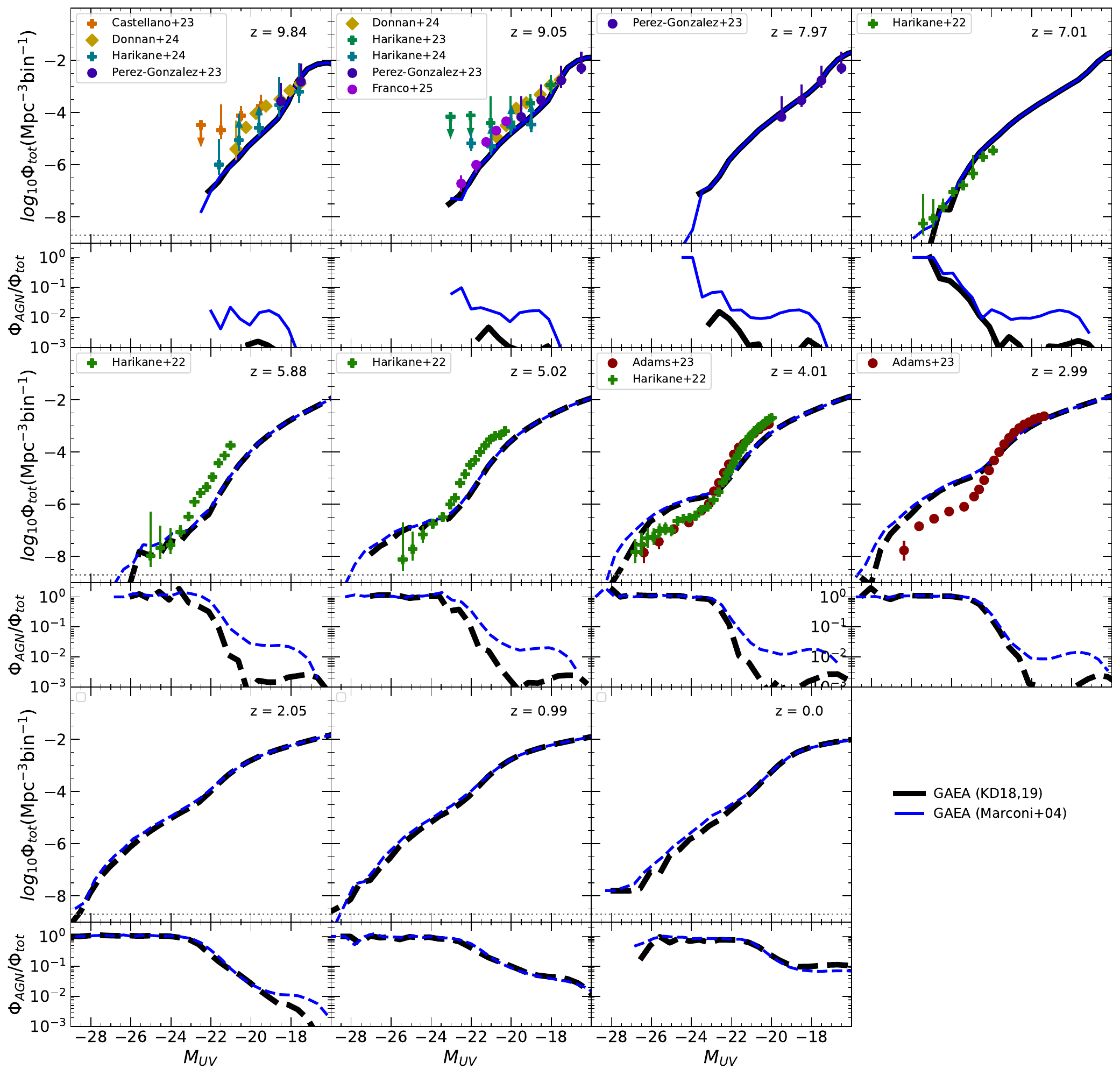}
	\caption{Total UVLF predicted by GAEA. Black lines are computed using the model from \cite{kubota_done2018,kubota_done2019}, also shown in \autoref{fig:GAEA_totUVLF_fiducial}, while blue lines are computed through the bolometric correction described in \cite{marconi2004}.}
	\label{fig:GAEA_totUVLF_comp_methods}
\end{figure*}

\autoref{fig:GAEA_totUVLF_comp_methods} shows a comparison of the total UVLF computed using the two methods described in Section \ref{sec:GAEA_fiducial_UVLF}. Both are in excellent agreement with each other, especially at $z \lesssim 2$. At $z \geq 3$, the models by \cite{kubota_done2018,kubota_done2019} predict a slightly larger contribution of AGN to the UV luminosity in the galaxy-dominated regime. 

\section{Black hole-stellar mass relation}\label{app:BH_stellar_mass_relation}
\begin{figure*}
	\centering
	\includegraphics[width=0.95\linewidth]{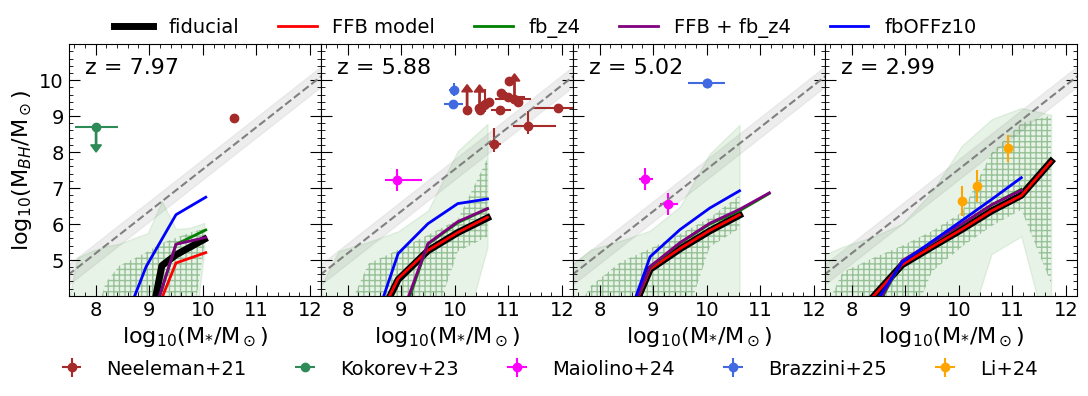}
	\caption{Black hole-stellar mass relation predicted by GAEA, along with observational data from \cite{neeleman2021}, \cite{kokorev2023}, \cite{lai2024}, \cite{maiolino2024}, and \cite{brazzini2025}. The solid lines show the median relation for each model variant considered, while the shaded areas correspond to the $1 \sigma$ (dark green) and $3 \sigma$ (light green) contours. The color code for the lines is the same as in \autoref{fig:GAEA_GSMF_variants}. The grey dashed line and shaded area correspond to the $z=0$ relation from \cite{kormendy_ho2013}.}
	\label{fig:GAEA_Mbh_Mstar_variants}
\end{figure*}

In GAEA, black holes accrete their mass from the surrounding gas, both during disc instability episodes and mergers \citep{fontanot2020b}. The implementation of a weaker stellar feedback might retain more gas inside the galaxy, which may contribute not only to an increase of star formation, but also to a faster black hole accretion. \autoref{fig:GAEA_MZR_variants} shows predictions from all models considered, along with observations from \cite{neeleman2021}, \cite{kokorev2023}, \cite{maiolino2024}, \cite{lai2024}, \cite{brazzini2025}, and the $z=0$ relation from \cite{kormendy_ho2013}. For the `no high-$z$ stellar feedback' model version, galaxies have more massive black holes than in other variants at fixed stellar mass and  in the redshift range $6 \leq z \leq 10$. When comparing model predictions with observational estimates, we find that all model variants, as well as our fiducial GAEA predictions, underpredict the estimated black hole masses in all redshift range considered. We underline that the statistical uncertainties related to such measurements, including the ones shown in \autoref{fig:GAEA_Mbh_Mstar_variants}, should also be convolved with systematic errors due to the methods adopted to estimate the black hole and stellar masses, that can reach values of the order of \SI{0.5}{\dex} \citep{brazzini2025}. Addressing such mismatch between predictions and estimates goes beyond the goals of our work.

\end{appendix}
\end{document}